\documentclass[useAMS,usenatbib]{mn2e}
\usepackage{epsf}
\usepackage{amsmath}

\newcommand{\hkpc}{h^{-1}{\rm kpc}}
\newcommand{\hmpc}{h^{-1}{\rm Mpc}}

\newcommand{\msun}{M_{\odot}}

\newcommand{\aap}{A\&A}
\newcommand{\apjs}{ApJS}
\newcommand{\apj}{ApJ}
\newcommand{\apjl}{ApJL}
\newcommand{\aj}{AJ}

\newcommand{\mnras}{MNRAS}

\newcommand{\hii}{\hbox{H\,{\sc II}}}
\newcommand{\heii}{\hbox{He\,{\sc II}}}
\newcommand{\taumax}{\tau_{\mathrm{max}}}
\newcommand{\nd}{n_{\mathrm{d}}}
\newcommand{\fJ}{f_{\mathcal{J}}}
\newcommand{\xhi}{x_{\mathrm{H\,I}}}
\newcommand{\ngrid}{n_{\mathrm{grid}}}
\newcommand{\tlc}{t_{\mathrm{LC}}}

\begin{document}
\onecolumn

\title{A New Moment Method for Continuum Radiative Transfer in
Cosmological Reionization}

\author[astronomy]{Kristian Finlator, Feryal \"{O}zel, \& Romeel Dav\'e}

\maketitle

 \begin{abstract}
We introduce a new code for computing time-dependent continuum
radiative transfer and non-equilibrium ionization states in static density 
fields with periodic boundaries.  Our code solves the moments of the 
radiative transfer equation, closed by an Eddingtion tensor computed
using a long characteristics method.  We show that pure (i.e.,
not source-centered) short characteristics and the optically-thin
approximation are inappropriate for computing Eddington factors for the
problem of cosmological reionization.  We evolve the non-equilibrium
ionization field via an efficient and accurate (errors $<1\%$) technique
that switches between fully implicit or explicit finite-differencing
depending on whether the local timescales are long or short compared
to the timestep.  We tailor our code for the problem of cosmological
reionization.  In tests, the code conserves photons, accurately treats
cosmological effects, and reproduces analytic Str\"omgren sphere 
solutions.  Its chief weakness is that the computation time for the long 
characteristics calculation scales relatively poorly compared to other 
techniques ($\tlc \propto N_{\rm{cells}}^{\sim1.5}$); however, we mitigate 
this by only recomputing the Eddington tensor when the radiation field 
changes substantially.  Our technique makes almost no physical 
approximations, so it provides a way to benchmark faster but more 
approximate techniques.  It can readily be extended to evolve multiple 
frequencies, though we do not do so here.  Finally, we note that our 
method is generally applicable to any problem involving the transfer of 
continuum radiation through a periodic volume.
\end{abstract}

 
\section{Introduction} \label{sec:intro}

The epoch of reionization is the current frontier in understanding
how galaxies form and evolve over cosmic time.  After the Universe
cooled sufficiently to recombine hydrogen atoms at redshift $z\approx
1088$~\citep{spe07}, the Universe was fully neutral.  Gravity grew
ever-denser structures that, at $z\sim 30-50$, were able to collapse into
stars and/or black holes.  The radiation emitted from these first objects
then began to re-ionize hydrogen.  By $z\sim 6$, hydrogen reionization
appears to be complete~\citep{fan07}, and the diffuse intergalactic medium
(IGM) has a neutral fraction of $\sim 10^{-4}$.  Understanding this
transition epoch is central to understanding the origin of galaxies and
the evolution of the IGM.  It is a major science driver for a host of
upcoming international telescope facilities, such as the {\it James Webb
Space Telescope} and the {\it Atacama Large Millimeter Array}.

Reionization involves a complex interplay between nonlinear growth
of structure, radiative cooling, star/black hole formation, chemical
enrichment, and photon transport.  Numerical simulations are required
to accurately model these highly nonlinear processes.  However, the large
dynamic range and complex physics involved make this an extraordinarily
challenging computational problem.  To obtain a full picture of
reionization in the context of currently-favored hierarchical structure
formation models, it is imperative that simulations include processes of
star formation, galaxy formation, and IGM evolution, along with feedback
processes that connect all three.  Cosmological hydrodynamic simulations
accounting for these processes are now achieving maturity, thanks
to improving algorithms and computing power.  However, the inclusion
of radiation transport complicates matters immensely.  A cosmological
radiative hydrodynamics code that can accurately evolve a representative
volume with sufficient dynamic range to study how galaxies reionize the
Universe would be a major development towards understanding reionization.
In this paper we provide a step towards that end by introducing a new
accurate moment-based method for calculating radiative transfer (RT) 
in a cosmological context.

Time-dependent radiative transfer is one of the most difficult components
to treat in any theoretical study of the reionization epoch owing to the
problem's well-known high dimensionality.  Consequently, over the past
decade, a number of approximate treatments have emerged that seek to
render it more tractable through well-motivated physical approximations.
The most flexible methods are the fully analytic treatments~\citep[for
example,][]{mad99,wyi03,fu04,ili05,kra06}.  These generally involve
assuming values for quantities such as the gas clumping factor and the
recombination rate that are averaged over all space or, in the case of
the excursion-set formalism~\citep{fu04}, over the volume of an ionized
region.  In exchange, they readily allow for broad surveys of parameter
space to be performed.

The next step in the direction of a full solution is taken by the
semi-numerical methods~\citep{cia00,mes07,gei08,cho08}, which combine
numerically-generated density fields with analytic treatments for
radiative transfer using techniques such as the excursion-set formalism
in order to account more realistically for source bias and the effects
of inhomogeneous density fields.  These treatments offer a dramatic
increase in realism over purely analytic calculations at modest additional
computational cost.  However, they have some difficulty accounting fully
for the consequences of inhomogeneous density fields such as shadowing
and the tendency for low-density regions to have a lower neutral fraction
during the later stages of reionization~\citep{cho08}.  These problems
arise from the need of semi-numerical models to make assumptions regarding
the shape of the ionized regions surrounding individual sources and the
nontrivial relationship between dark matter and gas densities in the
nonlinear regime.

Some of these difficulties are avoided in models that actually solve
the radiative transfer equation on numerically-generated density fields
but without fully accounting for radiative feedback on the sources
(\citealt{cia01,sok01,mel06,mcq07,ili07a}; see also \citealt{ili06} for a
very useful comparison of a number of techniques).  Nonetheless, obtaining
realistic baryonic density and emissivity fields in such contexts still
presents considerable challenges~\citep{mcq07}.  Additionally, while
parametrized treatments for radiative feedback have been introduced in
such models in order to study, for example, whether the photoevaporation
of minihaloes extends the epoch of reionization~\citep{ili07a,mcq07},
the simplified nature of these studies leaves their results open
to question~\citep{mes07}.  Hence, while each of these methods has
yielded an abundance of insight into reionization and warrant continued
development, the need is emerging for a complete solution to the radiative
transfer equation that is merged self-consistently with hydrodynamical
calculations.

A few fully radiative-hydrodynamic codes have been used to study
the reionization epoch~\citep{gne01,cen02,rij06}.  Unfortunately,
the techniques that these codes have introduced do not yet enjoy
widespread use owing to their high computational expense, even though
such studies are crucial for tuning the assumptions employed in more
simplified treatments and verifying their conclusions.  Moreover,
despite the enormous gain in realism that these codes have afforded the
reionization community, they too involve some physical approximations
such as the use of the ``optically thin variable Eddington tensor"
approximation~\citep{gne01} and a reduced~\citep{gne01} or an infinite
speed of light~\citep{cen02,rij06}.

In this work, we present a moment method solution to the radiative
transfer equation~\citep{aue70} and test it on static density fields.
Our technique is highly flexible, involves a minimum of physical
approximations, and can readily be combined with existing hydrodynamical
calculations.  It is similar to the method presented by~\citet{sto92},
but with several differences.  First, we optimize our code only for
cubical simulation volumes with periodic boundaries, as this is typical of
cosmological simulations.  Second, we derive our Eddington tensors from
a long characteristics (LC) calculation in order to minimize artifacts 
owing to poor angular and spatial resolution.  Finally, we include a 
treatment for nonequilibrium ionizations and account for the cosmological 
terms in the radiative transfer equation.  In a follow-up paper, we will 
present its implementation within a cosmological galaxy formation code.

We begin in Section~\ref{sec:rt} by casting the radiative transfer
equation into the form in which we solve it and summarizing our numerical
method.  In Section~\ref{sec:fedd}, we compare the performance of long
characteristics versus two other time-independent radiative transfer
techniques in order to select a method for deriving the Eddington tensor,
which we need in order to close our moment hierarchy.  After demonstrating
that long characteristics introduces the fewest unphysical artifacts,
we optimize it for computing reionization using a suite of realistic
albeit low-resolution integrations.  In Section~\ref{sec:nlte}, we
discuss our technique for evolving the nonequilibrium ionization field.
In Section~\ref{sec:full_alg}, we summarize our iterative scheme
for weaving these ingredients into a self-consistent calculation.
In Section~\ref{sec:tests}, we subject our code to a number of
standard tests.  Finally, we summarize our method and results in
Section~\ref{sec:summary}.

\section{Solving the Radiative Transfer Equation} \label{sec:rt}
We begin this section by writing down the RT equation in comoving 
coordinates including emission, absorption, and cosmological effects.
Next, we recast the RT equation in the form that our code computes
and discuss our treatment of the various terms.  Finally, we discuss 
our approach to solving these equations numerically.  

\subsection{The Moments of the Radiative Transfer Equation} \label{sec:rt-moments}
The radiative transfer equation in comoving coordinates is~\citep[for a
derivation, see][]{gne97}
\begin{equation} \label{eqn:rt}
\frac{1}{c}\frac{\partial \mathcal{N}_{\nu}(\hat{n})}{\partial t} + 
  \frac{\hat{n}}{a} \cdot \vec{\nabla}_c \mathcal{N}_{\nu}(\hat{n}) +
  {H}\left( 2 \mathcal{N}_{\nu} - \nu\frac{\partial \mathcal{N}_{\nu}}{\partial \nu} \right)
  =
  c \eta_{\nu} - c \chi_{\nu} \mathcal{N}_{\nu}(\hat{n}).
\end{equation}
Here, $\mathcal{N}_{\nu}(\hat{n})$ represents the number of 
photons with frequency between $\nu$ and $\nu+d\nu$ crossing an area $dA$ 
in the direction $\hat{n}$ into a solid angle $d\Omega$ during a time 
interval $dt$; $H$ is the Hubble constant; $a$ is the cosmological 
expansion factor; $\vec{\nabla}_c$ denotes a gradient in comoving 
coordinates, $\eta_{\nu}$ is the local number of photons emitted per 
unit time per unit solid angle with frequency between $\nu$ and 
$\nu+d\nu$; $c$ is the speed of light; and the absorption mean opacity 
$\chi_\nu=\sum_i n_i \sigma_{\nu,i}$ is the sum of the opacities due 
to the various absorbing species.  All emissivities and opacities are
taken as isotropic because Equation~\ref{eqn:rt} is written 
in the cosmological comoving frame.  Here and throughout our work, 
we compute the radiation field in terms of photon number densities 
rather than energy densities.  For simplicity of notation, we will 
generally omit the $\hat{n}$-dependence of $\mathcal{N}_\nu$ from now on.

The left hand side of Equation~\ref{eqn:rt} is the convective 
derivative of the photon phase space density but written in terms of 
$\mathcal{N}_\nu$.  The first two terms are the classical convective 
derivative modified to apply in comoving spatial coordinates, and the 
terms proportional to the Hubble constant account for, respectively, the 
dilution of $\mathcal{N}_\nu$ and redshifting of the photon frequencies 
owing to cosmological expansion.  The terms on the right hand side 
account for photon emission and absorption, respectively.

By integrating over the frequency range $(\nu_1,\nu_2)$, we recast 
Equation~\ref{eqn:rt} in a form appropriate for a multigroup method:
\begin{equation} 
\label{eqn:rt-multigroup}
\frac{1}{c}\frac{\partial \mathcal{N}}{\partial t} + 
  \frac{\hat{n}}{a} \cdot \vec{\nabla}_c \mathcal{N}
  = 
  \eta - (\chi_H + \chi_{\rm{abs}})\mathcal{N}
\end{equation}
Here, we have defined the photon number density $\mathcal{N}$, the 
emissivity $\eta$, the cosmological opacity $\chi_H$, 
the spectral slope $\langle \nu \frac{\partial }{\partial\nu}\rangle$, 
and the absorption mean opacity $\chi_{\rm{abs}}$ as follows:
\begin{eqnarray}
\mathcal{N} & \equiv & \int^{\nu_2}_{\nu_1} \mathcal{N}_{\nu} d\nu \\
\eta & \equiv & \int^{\nu_2}_{\nu_1} \eta_{\nu} d\nu\\
\chi_H & \equiv & \frac{H}{c}
  \left(2-\langle\nu\frac{\partial}{\partial\nu}\rangle\right)
\label{eqn:chi_H}\\
\langle\nu\frac{\partial}{\partial\nu}\rangle & \equiv &
  \int^{\nu_2}_{\nu_1} \nu \frac{\partial \mathcal{N}_{\nu}}{\partial \nu} d\nu /
  \int^{\nu_2}_{\nu_1} \mathcal{N}_{\nu} d\nu  
\label{eqn:spec_slope} \\
\chi_{\rm{abs}} & \equiv &  
  \int^{\nu_2}_{\nu_1} \chi_{\nu} \mathcal{N}_{\nu} d\nu / 
  \int^{\nu_2}_{\nu_1} \mathcal{N}_{\nu} d\nu. \label{eqn:chi_abs}
\end{eqnarray}
Equation~\ref{eqn:rt-multigroup} is equivalent to an integral of
Equation~\ref{eqn:rt} over frequency as long as the frequency-averaged
cosmological and absorption mean opacities $\chi_H$ and 
$\chi_{\rm{abs}}$ can be determined consistently.  Both depend on
the slope of the spectrum at each frequency bin.  The dependence of the
absorption mean opacity $\chi_{\rm{abs}}$ is clear, and the dependence of 
the cosmological opacity $\chi_H$ can be made more intuitive by noting 
that, for a power-law spectrum $\mathcal{N}_{\nu} \propto \nu^{-\alpha}$, 
the spectral slope $\langle\nu\frac{\partial}{\partial\nu}\rangle$ is given 
by $-\alpha$.  In this case, the cosmological opacity $\chi_H$ reduces to 
$\frac{H}{c}(2+\alpha)$.  This term is generically quite small in the 
problem of cosmological reionization: For star-forming galaxies and active
galactic nuclei, the slope of the ultraviolet continuum generally falls 
within the range $\alpha \sim 0.3$--$5$, so during the reionization 
epoch $\chi_{H} \sim 0.1$--$3\times10^{-26}$cm$^{-1}$.  By contrast, 
even for a neutral hydrogen fraction of $10^{-3}$ at 
$z=6$~\citep{fan06}, the opacity at 912\AA\ at the mean density 
is around $3$--$4\times10^{-25}$cm$^{-1}$.  This is simply a 
statement that, throughout the reionization epoch, ionizing 
photons tend to be absorbed long before they can be diluted or 
redshifted by the Hubble flow.  For this reason, we bring the cosmological
term to the right-hand side of Equation~\ref{eqn:rt-multigroup}.  In
multifrequency computations we will allow the spectral slope $\alpha$ to 
vary self-consistently with frequency by using the value from the previous
timestep.  However, in the present work, we neglect it entirely as we 
consider only monochromatic problems.  From now on, we use the opacity 
$\chi$ to refer to the sum of the cosmological and absorption mean 
opacities: $\chi \equiv \chi_H + \chi_{\rm{abs}}$.

The first two angle moments of Equation~\ref{eqn:rt-multigroup} are:
\begin{eqnarray} 
\label{eqn:rt-moments-E}
\frac{\partial \mathcal{J}}{\partial t} & = &
  -\frac{1}{a} \vec{\nabla}_c \cdot \vec{\mathcal{F}} + 4\pi\eta - c \chi \mathcal{J} \\
\label{eqn:rt-moments-F}
\frac{\partial \vec{\mathcal{F}}}{\partial t} & = &
  -\frac{c}{a} \vec{\nabla}_c \cdot (c\mathbf{f}\mathcal{J}) - c \chi \vec{\mathcal{F}} \\
\label{eqn:rt-moments-f}
\mathbf{f} & \equiv & \frac{\int \mathcal{N} \hat{n}\hat{n} d\Omega}{\int \mathcal{N} d\Omega}.
\end{eqnarray}
The zeroth and first moments $\mathcal{J}$ and $\vec{\mathcal{F}}$ are the 
angle-averaged mean number density (hereafter ``number density") and flux of 
photons in a frequency bin, respectively, and the Eddington tensor 
$\mathbf{f}$ is used to close the moment hierarchy; we will discuss how we 
obtain the Eddington tensor in Section~\ref{sec:fedd}.

\begin{figure}
\setlength{\epsfxsize}{0.5\textwidth}
\centerline{\epsfbox{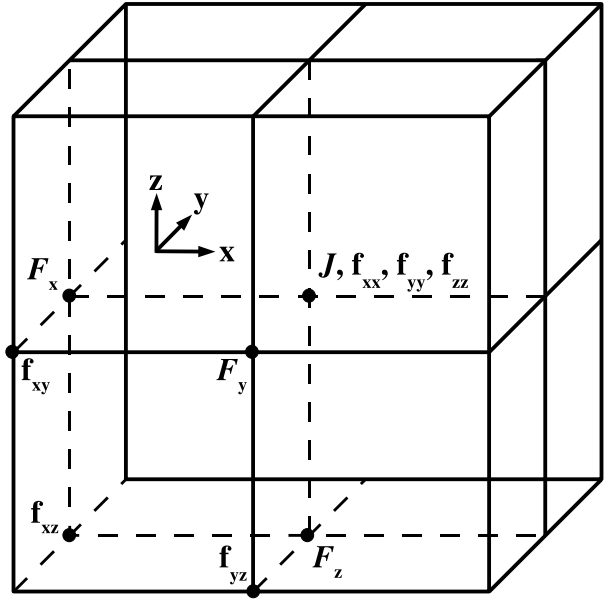}}
\vskip -0.0in
\caption{The centering scheme for our radiative variables.}
\label{fig:centering}
\end{figure}

When constructing a solution to equations~\ref{eqn:rt-moments-E}
and~\ref{eqn:rt-moments-F}, it is important to center the
photon number density, flux, and the components of the Eddington
tensor about each cell spatially in such a way that the resulting
finite-difference approximation for the number density at the updated
time (Equation~\ref{eqn:rt-moments-tstep-E}) is cell-centered.  This is
important not only to improve the solution's accuracy in space, but
also because improper centering can lead to spurious anisotropies in the
radiation field or even prevent the solution from converging altogether.
Accordingly, we center the variables as follows: $\mathcal{J}$
and diagonal components of $\mathbf{f}$ are positioned at the cell
center, components of $\vec{\mathcal{F}}$ are stored at cell faces,
and off-diagonal elements of $\mathbf{f}$ are stored at cell edges
(see Figure~\ref{fig:centering}).

One ambiguity remains regarding the treatment of off-diagonal elements
of $\mathbf{f}$ such as $\mathbf{f_{xy}} \equiv \int \mathcal{N}
\hat{x}\hat{y}d\Omega / \int \mathcal{N} d\Omega$.  Proper centering
requires that these factors always enter into the partial derivatives
of Equation~\ref{eqn:soln} via a spatially-averaged product with the
photon number density $\mathcal{J}$.  For example, the finite-difference
approximation for $\mathcal{J}^{n+1}(i,j,k)$ includes, among other
things, the spatial average of the product $\mathbf{f_{xy}}\mathcal{J}$
over the cells $(i,j,k), (i-1,j,k), (i,j-1,k),$ and $(i-1,j-1,k)$.
In such instances, it is possible to use either the ``average of the
products", $\langle \mathbf{f} \mathcal{J}\rangle$ or the ``product of
the averages", $\langle\mathbf{f}\rangle \langle \mathcal{J} \rangle$.
Testing suggests that the differences between the two options are small,
but inspection shows that the ``average of the products" option gives
rise to significant cancelling of terms within the full finite-difference
expression and hence could lead to less smooth solutions.  For this reason,
we prefer the ``product of the averages" approach.

\subsection{Solving the Radiation Transfer Equation}\label{sec:rt-soln}
We now discuss our technique for integrating
Equations~\ref{eqn:rt-moments-E} and~\ref{eqn:rt-moments-F}
numerically.  These equations can be quite stiff when applied to
the problem of cosmological reionization, hence we use an implicit
finite-differencing scheme in time.  In other words, we use the
photon number density and flux at the updated time on the right hand
side of Equations~\ref{eqn:rt-moments-E} and~\ref{eqn:rt-moments-F}.
SMN92 solved the resulting system using the ``Automatic Flux-Limiting"
prescription of~\citet{mih82}.  This scheme proceeds by integrating
Equation~\ref{eqn:rt-moments-F} over one timestep analytically before
plugging the result into an implicitly finite-differenced form of
Equation~\ref{eqn:rt-moments-E}.  However,~\citet{hay03} found that
the results from this technique do not differ significantly from the
results of simply finite-differencing Equation~\ref{eqn:rt-moments-F}.
Our own testing also indicates that the latter technique produces accurate
results, hence in our code the updated flux $\vec{\mathcal{F}}^{n+1}$
and number density $\mathcal{J}^{n+1}$ relate to the previous values
$\vec{\mathcal{F}}^{n}$ and $\mathcal{J}^{n}$ as follows:
\begin{eqnarray} 
\label{eqn:rt-moments-tstep-E}
\mathcal{J}^{n+1} & = & 
  \frac{1}{1+x^{n+1}} \left[ 
    \mathcal{J}^n + 4\pi\eta\Delta t - 
    \Delta t \frac{\vec{\nabla}_c}{a} \cdot \left(
      \frac{\vec{\mathcal{F}}^n}{1+x^{n+1}}
    \right) +
    c^2 \Delta t^2 \frac{\vec{\nabla}_c}{a} \cdot \left[
      \frac{1}{1+x^{n+1}} \frac{\vec{\nabla}_c}{a} \cdot (\mathbf{f}\mathcal{J}^{n+1})
    \right]
  \right] \\
\label{eqn:rt-moments-tstep-F}
\vec{\mathcal{F}}^{n+1} & = & 
  \frac{1}{1+x^{n+1}} \left[ \vec{\mathcal{F}}^{n} - 
  c^2 \Delta t \frac{\vec{\nabla}_c}{a} \cdot (\mathbf{f} \mathcal{J}^{n+1}) \right] \\
\nonumber
x^{n+1} & \equiv & c \chi^{n+1} \Delta t
\end{eqnarray}

These equations can be combined and rearranged into the form
\begin{equation}
\label{eqn:soln}
\mathbf{A} \cdot \vec{\mathcal{J}}^{n+1} = \vec{b},
\end{equation}
where the vector notation indicates that we are solving a coupled
system of algebraic equations with dimension equal to the number of
computational cells.  The update matrix $\mathbf{A}$ is a function 
of $\vec{\mathcal{J}}^n$ and $\vec{\mathcal{F}}^n$ but not $\vec{\mathcal{J}}^{n+1}$, hence the photon
number densities may be updated via a simple matrix inversion.  
The photon number density in each cell couples only to 18 of its 
neighboring cells.  Consequently, only 19 elements in each row of
$\mathbf{A}$ are nonzero, making it a sparse matrix.

Linear systems of equations for which the coefficient matrix is sparse but
not both symmetric and positive definite can be solved via the biconjugate
gradient method.  We have found that the preconditioned biconjugate
gradient routine \emph{linbcg} in \emph{Numerical Recipes}~\citep{nr}
solves the problem rapidly.  We use the diagonal of $\mathbf{A}$ as the
preconditioner and halt the iteration when the residual $|\mathbf{A} \cdot
\vec{\mathcal{J}}^{n+1} - \vec{b}|/|\vec{b}|$ is less than $10^{-6}$.

We generalize this technique to multifrequency problems by solving
Equations~\ref{eqn:rt-moments-tstep-E} and~\ref{eqn:rt-moments-tstep-F}
independently for a number of multigroup frequency bins.  We compute the 
Eddington tensor $\mathbf{f}$, absorption mean opacity, and cosmological 
opacity fields separately for each frequency bin.  When evaluating
Equations~\ref{eqn:chi_H}--\ref{eqn:chi_abs}, we use the spectrum
$\mathcal{J}(\nu)$ from the previous timestep.

\section{Computing the Eddington Factors} \label{sec:fedd}
As is well-known, the primary difficulty in solving the radiative transfer
equation lies in its high dimensionality.  Treating the moments of the 
equation does not suppress this dimensionality unless the photon mean free 
path is short compared to all length scales of interest, in which case one 
can close the moment hierarchy with analytical 
flux-limiters~\citep[e.g.,][]{hay06}.  This is not the case in the 
problem of cosmological reionization, hence an accurate solution depends 
critically on an accurate derivation of the Eddington tensor 
$\mathbf{f}$.  This in turn requires knowing how the photon density 
$\mathcal{N}$ varies as a function of direction $\hat{n}$ (see 
Equation~\ref{eqn:rt-moments-f}).  A fully consistent treatment would 
obtain $\mathcal{N}(\hat{n})$ via a time-dependent integration of the 
radiative transfer equation; however, if this were easily done then of 
course the entire problem would already be solved.  Here, we derive the 
Eddington tensor $\mathbf{f}$ from a time-independent formal solution to
the radiative transfer equation (that is, a solution in which
$\partial \mathcal{N}_{\nu} / \partial t = 0$; see also~\citealt{aue70};
SMN92).  We note that this is our only approximation.

Previous efforts have approached this problem through
computationally \emph{efficient} techniques such as short
characteristics~\citep{sto92,hay03} or the optically thin
approximation~\citep{gne01}.  While these techniques have their strengths
and have led to a great deal of insight into reionization, the associated
compromises in accuracy are poorly-understood.  For this reason,
we undertake an \emph{accurate} calculation of the Eddington tensor,
even though this degrades our computational efficiency, in order to
study its impact on cosmological problems.  In this section, we begin
by reviewing the long characteristics (LC) technique for computing
the Eddington tensor $\mathbf{f}$.  We then compare it to SC and the
optically thin approximation in order to highlight the strengths of LC.
Finally, we optimize LC for calculations of cosmological reionization.

\subsection{Long Characteristics}
\begin{figure}
\setlength{\epsfxsize}{0.5\textwidth}
\centerline{\epsfbox{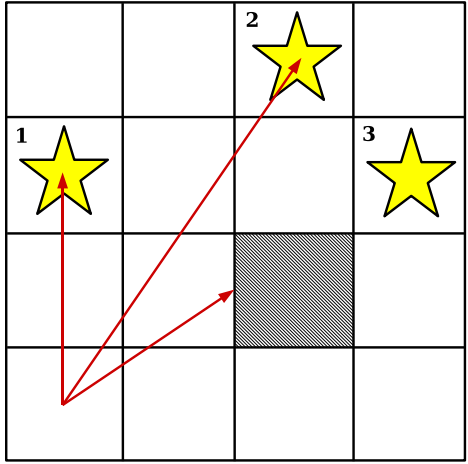}}
\vskip -0.0in
\caption{A sketch of our long characteristics method in two
dimensions.  Long characteristics from the bottom-left cell to sources 1 
and 2 are shown; the line integral to source 3 is halted when it encounters
an intervening optically thick cell.}
\label{fig:lc}
\end{figure}
The LC approach to computing $\mathcal{N}(\hat{n})$ at a target cell 
consists of integrating the (time-independent) radiative transfer equation 
along characteristics that run in the directions $\hat{n}_i$ from the 
target cell to the source cells $i$.  The photon density at the target 
cell is then given by the sum 
\begin{equation}
\mathcal{N}(\hat{n}) = \sum_i\frac{\eta_{i}}{\chi_i} e^{-\tau_i}\delta(\hat{n}-\hat{n}_i),
\end{equation}
where the index $i$ runs over all source cells and $\tau_i$ is 
the total optical depth between the target cell and the source cell $i$. 
The problem of computing $\mathcal{N}(\hat{n})$ at a target cell hence reduces 
to one of determining the total optical depth $\tau_i$ to each source cell. 
Here, we outline our approach to computing $\tau_i$ (see also 
Figure~\ref{fig:lc}).  We note that our treatment is similar to the 
ray-tracing technique of~\citet{abe99}.

Consider the contribution to the Eddington tensor in a target cell whose
center is located at $\vec{r}_T\equiv(x_T,y_T,z_T)$ owing to a source 
cell $i$ whose center is located at $\vec{r}_i\equiv(x_i,y_i,z_i)$.  We 
compute the optical depth between the two cells by integrating along a 
ray that points in the direction 
$\hat{n} = (n_x,n_y,n_z) = (\vec{r}_i - \vec{r}_T)/|\vec{r}_i - \vec{r}_T|$.
Starting at $\vec{r}_T$, the distance in the direction $\hat{n}$ to the 
nearest $y$-$z$ plane $\Delta r_x$ is given by
\begin{equation}
\Delta r_x = \left[\frac{\Delta x}{2} - \mathrm{sign}(n_x)(x-x_c)\right] 
  \sqrt{1 + (\frac{n_y}{n_x})^2 + (\frac{n_z}{n_x})^2},
\end{equation}
where sign($x$) equals $-1$ or $+1$ if $x$ is negative or positive,
respectively.  Analogous relations exist for the distance to the next 
$x$-$z$ boundary $\Delta r_y$ and the next $x$-$y$ boundary $\Delta r_z$.  
We determine which cell the ray enters next by evaluating which of the 
three distances is shortest; for example, if $\Delta r_x < \Delta r_y$
and $\Delta r_x < \Delta r_z$, then the ray will next encounter a $y$-$z$ 
plane.  We then add the contribution $\chi \Delta r$ to the optical depth 
from the cell that the ray just traversed assuming that $\chi$ is uniform 
throughout the cell.  Repeating this procedure, we continue adding 
contributions to the line integral until either the ray enters the source
cell or the accumulated optical depth exceeds a maximum value $\taumax$ that 
will be determined from convergence testing.  If the accumulated optical 
depth exceeds $\taumax$ before the ray enters the source cell $i$, then we 
consider the source to be completely obscured and halt the line integral.  
Otherwise, we add the contribution of the source cell to the Eddington 
tensor at the target cell and proceed to the next source cell.

If the cell is itself a source, then we account for the contribution of its
self-illumination to its Eddington tensor $\mathbf{f}$ by adding
$(4\pi \mathcal{N}_0/3)$\textbf{1} to the numerator and $4\pi \mathcal{N}_0$ to the
denominator in Equation~\ref{eqn:rt-moments-f}, where \textbf{1} indicates 
the unit tensor and $\mathcal{N}_0$ is given by
\begin{eqnarray*}
\mathcal{N}_0 & = & \frac{\eta}{\chi}(1-e^{-\chi r_*}) \\
r_* & = & \left( \frac{\Delta x \Delta y \Delta
z}{\frac{4}{3}\pi}\right)^{\frac{1}{3}}.
\end{eqnarray*}

\subsection{Comparison of Techniques}\label{sec:fedd-comp}
\subsubsection{Short Characteristics}
The method of short characteristics (SC) has been introduced 
elsewhere~\citep{kun88,sto92} and we will only summarize it here.  The SC 
approach involves solving the radiative transfer equation within each cell 
along characteristics that run from the cell's faces to its center
using boundary conditions at the cell faces that are obtained through 
interpolation.  Given the boundary conditions on the computational volume, 
one simply marches downstream from one side of the volume to the other so that, 
at each cell, the upstream boundary conditions are always known.  As long as 
the opacity and emissivity do not change dramatically on the scale of a 
computational cell and some spatial diffusion of the radiation field is 
acceptable, SC is an efficient technique for computing a time-independent 
formal solution to the RT equation (or even a time-dependent one; 
see~\citet{hay03}): In three dimensions, the computation time scales with 
the number of cells on each side of the computational grid $\ngrid$ as 
$O(\ngrid^3)$.  Unfortunately, if the emissivity or opacity varies 
significantly on the scale of the computational grid, then the 
interpolations can give rise to dramatic numerical artifacts in the spatial 
distribution of the photon number density.  In the problem of cosmological 
reionization, sources are generally pointlike and there are sharp 
transitions between optically thick and thin regions (for example, at 
ionization fronts), hence such artifacts are expected.  Additionally, given 
that the emissivity is highly concentrated spatially, the number of angles 
that must be sampled in order to yield smooth ionization fronts can scale 
as poorly as $O(\ngrid^2)$, yielding a less-favorable overall scaling of 
$O(\ngrid^5)$ (see, for example,~\citealt{raz99}).

\begin{figure}
\setlength{\epsfxsize}{0.5\textwidth}
\centerline{\epsfbox{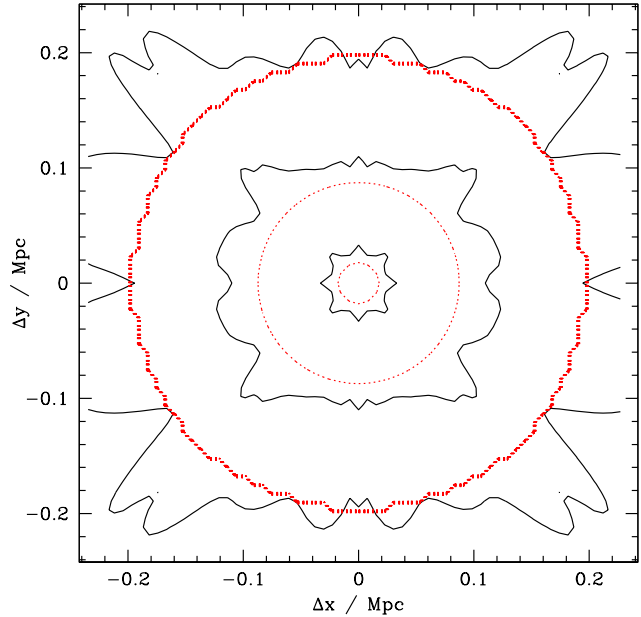}}
\vskip -0.0in
\caption{Contours in photon number density versus position in a plane that 
contains the source as calculated using SC (black solid) and LC (red dotted).
The strong anisotropies resulting from the interpolations that are inherent 
to SC compare unfavorably to LC, which does not involve interpolations.}
\label{fig:test_sc}
\end{figure}

The tendency for collimated radiation fields to diffuse numerically in 
SC owing to the interpolated boundary conditions at each cell has been 
discussed elsewhere~\citep[for example,][]{kun88}.  However, the 
anisotropies that can also result from these interpolations have not 
received much attention.  In order to demonstrate this problem, we 
consider the idealized case of a galaxy consisting of $10^8 \msun$ of 
young stars at $z=20$ with an ionizing escape fraction of 10\%.  We 
locate the galaxy at the center of a homogeneous region 0.5 (proper) Mpc 
on a side in which pure hydrogen of total number density 
$1.5\times10^{-3}$ cm$^{-3}$ and temperature $10^4$ K is ionized to a 
neutral fraction of $10^{-3}$.  We solve the time-independent radiative 
transfer equation for this scenario using both SC and LC.  In the SC 
integration we sample the unit sphere with 320 uniformly-distributed unit 
vectors.  In both cases we use a grid resolution of $64^3$ cells.

In Figure~\ref{fig:test_sc} we show contours of photon number density 
versus position in a plane that contains the source.  The interpolations that
are inherent to SC give rise to strong anisotropies in the number density
field owing to the discontinuous emissivity field at the source.  These can 
be alleviated (but not eliminated) by significantly increasing spatial and 
angular resolution, but at the cost of increased computation time.  Not
surprisingly, tests indicate that using SC-derived Eddington factors to
solve the Str\"omgren Sphere problem within our full time-dependent moments 
method gives rise to unacceptable anisotropies in the shape of the 
ionization front.  By contrast, LC yields smooth mean intensity contours 
because it does not involve any interpolations.

We have compared the performance of LC only to simple (that is, not
source-centered) SC.  A source-centered SC 
method~\citep[for example,][]{mel06} would probably perform differently 
although the required interpolations are still likely to lead to numerical 
artifacts.  However, we have opted not to consider source-centered SC here 
because it is not obvious that the computation time for this variant scales 
more favorably with spatial resolution than it does for LC.

\subsubsection{Optically Thin Approximation}

\begin{figure}
\setlength{\epsfxsize}{0.5\textwidth}
\centerline{\epsfbox{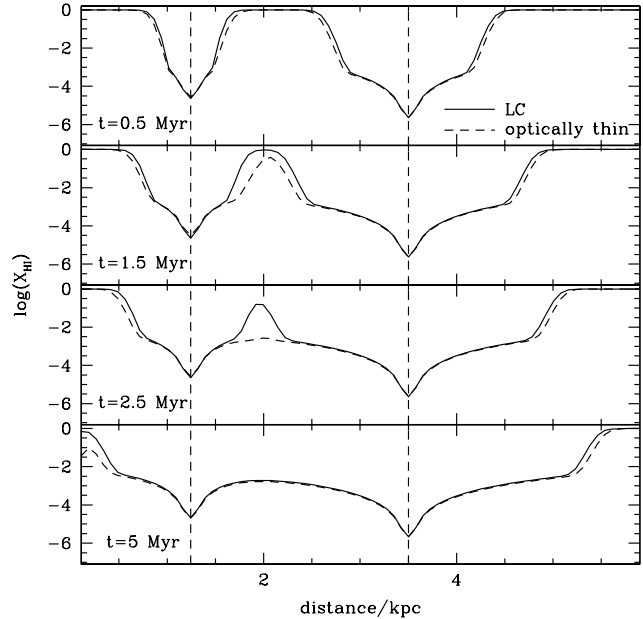}}
\vskip -0.0in
\caption{Neutral fraction as a function of position along a line connecting
a bright source and a faint source at three different times.  The solid and
dashed curves show the results obtained when computing the Eddington tensor
through LC and the optically thin approximation, respectively.  The vertical
dashed lines indicate the positions of the sources.  The optically 
thin approximation does well at early and late times, but the individual 
\hii\ regions overlap too quickly and the ionization front from the 
faint source is too extended during the overlap phase.}
\label{fig:test_otvet2}
\end{figure}

The optically thin approximation~\citep{gne01} involves evaluating
Equation~\ref{eqn:rt-moments-f} via a time-independent formal solution
to the radiative transfer equation in which the opacity is neglected.
Specifically, $\mathcal{N}$ is calculated from the sum total of all photons 
emitted in the volume, with no attenuation.  This approach conserves 
photons, does not suffer from some of the grid-induced artifacts that can 
occur in the SC approach, and yields results that are qualitatively 
reasonable.  Moreover, the computation time scales as $O(N^3)$.  These 
desirable characteristics motivate us to evaluate the benefits of 
accurately accounting for the optical depth, and the most direct way to 
do so is simply to compare the results of using optically thin versus 
accurate Eddington tensors within our moment method.  Our LC code is 
well-suited for performing such a comparison.  For this reason, we revisit 
the problem of multiple sources embedded in an initially optically thick 
medium.  In the optically thin approximation, a source can affect the 
Eddington tensor in the vicinity of a neighboring source even before their 
respective \hii\ regions have overlapped.  This can potentially lead to 
errors in the shape of the resulting ionization fronts even before they 
overlap (see also~\citealt{gne01}).

We consider the problem of a bright source with a monochromatic 
ionizing luminosity of $5\times10^{48}$ s$^{-1}$ located 2.2~kpc from 
a faint source whose luminosity is  $5\times10^{47}$ s$^{-1}$.  Both 
sources are embedded in a homogeneous medium of pure hydrogen with number 
density $10^{-3}$ cm$^{-3}$ and temperature $10^4$ K.  The medium is 
initially entirely neutral.  In this arrangement, the flux from the bright 
source dominates that of the faint source at a distance from the faint 
source equal to one quarter of the faint source's Str\"omgren radius.  
We evolve this system with Eddington tensors obtained from LC and the 
optically thin approximation until the sources' \hii\ regions overlap at 
$t\approx 5$ Myr.  We use a grid of $80^3$ computational cells and disable 
periodic boundary conditions.

In Figure~\ref{fig:test_otvet2}, we compare the resulting neutral 
fractions along the line passing through the source centers before, 
during, and after overlap.  At $t=0.5$ Myr, the two \hii\ regions 
are evolving approximately correctly in the optically thin approximation
although there is a suggestion that photons stream too rapidly along the
direction connecting the two sources.  By $t=1.5$ Myr, there is a noticeable
tendency for the ionization fronts to advance too rapidly between the two 
sources in the optically thin case.  This tendency oversuppresses the neutral 
fraction in this region.  By $t=2.5$ Myr, it is clear that overlap has 
occurred too soon in the optically thin approximation.  Finally, after overlap 
has occurred ($t=5$ Myr) the optically thin approximation performs well again 
because, at least within the ionized region, it is no longer a strong 
approximation.

\begin{figure*}
\setlength{\epsfxsize}{1.0\textwidth}
\centerline{\epsfbox{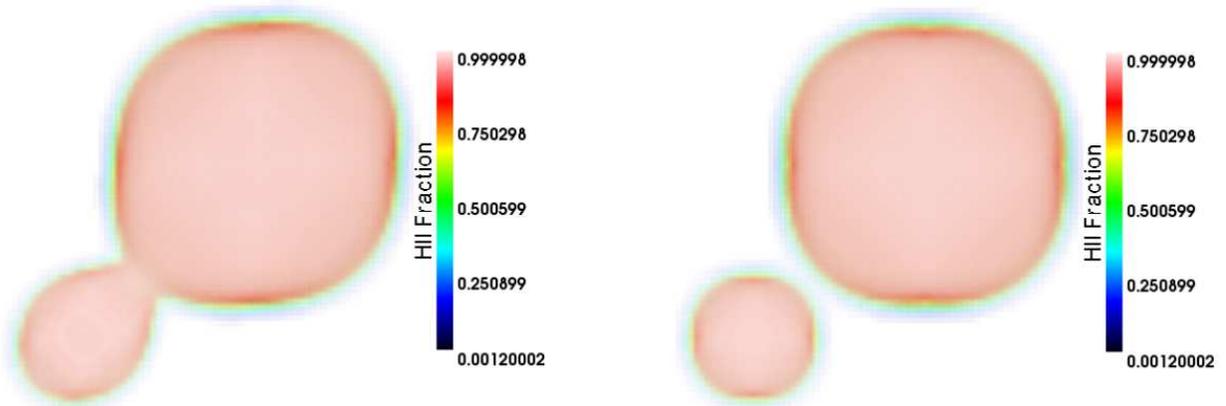}}
\vskip -0.0in
\caption{Ionized hydrogen fraction as a function of position for the test
in Figure~\ref{fig:test_otvet2} at $t=2.1$ Myr.  The left and right panels
show how the \hii\ regions appear when we compute the Eddington tensors using 
the optically thin approximation and long characteristics, respectively.  
The smaller \hii\ region is dramatically elongated in the optically thin 
approximation.  These figures were produced using {\sc ifrit}.}
\label{fig:test_otvet2_snaps}
\end{figure*}

In Figure~\ref{fig:test_otvet2_snaps}, we compare the morphologies of the 
\hii\ regions at $t=2.1$ Myr.  The left and right panels show the results 
of using the optically thin approximation and long characteristics, 
respectively.  Looking at the right panel first, we see that the \hii\ 
regions show slight departures from spherical symmetry even with accurate 
Eddington tensors, appearing slightly boxy in this projection owing to low 
spatial resolution.  In addition to this asymmetry, however, the smaller 
\hii\ region appears dramatically elongated prior to overlap when we use
the optically thin approximation to compute the Eddington tensors.  In 
fact, the optically thin approximation causes both \hii\ regions to elongate 
along the axis that connects the two sources, leading to the early overlap 
seen in Figure~\ref{fig:test_otvet2}.

What effect do morphological errors and a tendency towards early overlap 
have on galaxy evolution during the recombination epoch? 
Figures~\ref{fig:test_otvet2} and~\ref{fig:test_otvet2_snaps} suggest that 
the errors will be small on scales that are larger than the largest ionized 
regions at any given time.  The fact that moment methods automatically 
conserve photons irrespective of the Eddington tensors reinforces this view.  
However, at smaller scales it is possible that galaxy evolution in satellite 
halos will be oversuppressed, especially in regions between larger halos.  
Such an error could in turn lead to an underestimate of the number density 
of smaller ionized regions, which may dominate the photon budget of 
overdense regions~\citep{ili07b}.  The only way to settle this question 
will be through full-scale simulations of reionization in which the two 
techniques can be compared.

\subsection{Optimizing the Long-Characteristics Calculation} 
\label{sec:fedd-lcopt}

Our LC calculation of the Eddington tensors is time consuming.  To optimize 
the calculation, we introduce some numerical approximations.  Because these 
approximations are numerical and not physical, it is possible to rigorously 
assess the errors introduced through convergence tests.  The key optimization 
is that we only recompute the Eddington tensor field when the radiation field 
has changed significantly, and not at every timestep of the moment solver.  
Since in typical cosmological situations the radiation field evolves 
relatively slowly over much of the volume, this results in many fewer LC 
calculations.  In this section we discuss our optimizations and quantify the 
errors.  

The first problem that we must address in optimizing our LC technique
for cosmological reionization is the way that we smooth the Eddington
tensor field.  Smoothing is necessary because discontinuities in the
opacity field (for example, around isolated sources or at ionization
fronts) give rise to discontinuities in the Eddington tensor field, which
in turn imprint numerical artifacts onto the morphology of the radiation
and ionization fields.  Moreover, in extreme cases, discontinuities in
the Eddington tensors can prevent the code from converging altogether
(as also noted by~\citealt{raz99}).  While the optimal solution
to these problems would be to enforce a spatial resolution in which no
cell is optically thick, this condition is computationally prohibitive.
We have found that smoothing the Eddington tensor field with a 27-cell
tophat filter largely removes the numerical effects (see, however,
Figure~\ref{fig:test_otvet2_snaps}).  This smoothing does not degrade the
quality of our solution for two reasons: (1) it does not impact photon
conservation, and (2) it does not reduce the solution's spatial resolution
because the moment method is already second-order in space; that is,
we smooth over the same spatial scales to compute the spatial derivatives.

Next, we turn to the choice of numerical parameters.  Our implementation
of LC introduces three parameters that we optimize through convergence
tests: (1) The maximum optical depth from a target cell to a source cell
$\taumax$ beyond which we consider the source to be obscured and halt
its LC line integral; (2) the depth $\nd$ of periodic replicas that
we use to mimic periodic boundaries; and (3) the minimum fractional 
change in photon number density $\fJ$ required to trigger an update to 
a cell's Eddington tensor.  Generally, we will require a 10\% median 
accuracy on the photon number density $\mathcal{J}$.

Our convergence tests involve computing the reionization of a static
cosmological density and emissivity field.  We obtain this field by
dividing the gas and stellar densities from the $z=9$ snapshot of an
$8\hmpc$ cosmological volume (the w8n256vzw simulation of~\citealt{opp06})
onto a $16^3$ grid.  We divide the total mass associated with SPH
particles that lie near cell boundaries between the cells by summing
incomplete gamma functions to their equivalent Plummer SPH smoothing
kernels.  We assume that all gas is completely neutral at $z=9$.
We compute the cell emissivities by convolving the stellar populations
in each cell with the~\citet{bc03} stellar population synthesis models
and assuming a 10\% escape fraction for ionizing photons.  During the
integration, we account for cosmological expansion by assuming ($\Omega,
\Lambda, H_0$) = (0.3, 0.7, 70).  With this setup, we find that the
volume-averaged neutral fraction drops to roughly $\xhi=10^{-3}$ at
$z=6$ (Figure~\ref{fig:cosmo_x_vs_t}), in good agreement with available
constraints~\citep{fan06}.

\subsubsection{Optimizing the Maximum Optical Depth}
\label{sec:fedd-lcopt-taumax}
\begin{figure}
\setlength{\epsfxsize}{0.5\textwidth}
\centerline{\epsfbox{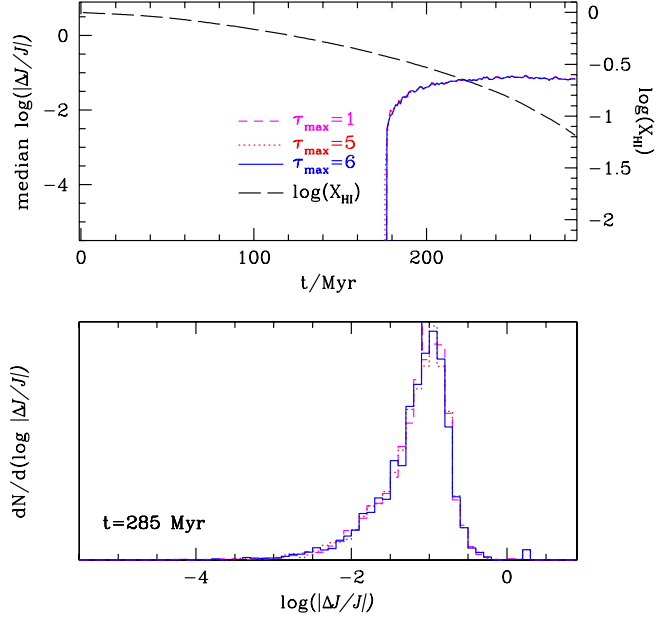}}
\vskip -0.0in
\caption{Convergence-testing the parameter that controls the optical depth
at which LC line integrals are terminated, $\taumax$, by comparing fractional 
errors in the local photon number density $\mathcal{J}$ as a function of time.  
We compute reionization using $\taumax=1$ (short-dashed magenta), $\taumax=5$ 
(dotted red) and $\taumax=6$, (solid blue); we obtain the ``converged" answer 
by assuming $\taumax=1000$.  The top panel shows the median local error in
$\mathcal{J}$ as a function of time while the bottom panel shows the full 
distribution of local fractional errors at $t=285$ ($z=6.5$, $\xhi = 0.07$).  
The vertical lines at the top of the bottom panel indicate the medians.  
Here and in Figures~\ref{fig:test_nd} and~\ref{fig:test_fJ}, we consider 
only cells with comoving photon number density 
$\mathcal{J} > 1\times10^{-10}$ cm$^{-3}$ in order to eliminate 
cells where $\mathcal{J}$ is dominated by roundoff error.  Choosing 
$\taumax=6$ yields median fractional accuracy errors $\leq 10\%$ at all 
times.}
\label{fig:test_taumax}
\end{figure}

We look for an optimal maximum optical depth $\taumax$ beyond which
the effect of terminating the LC line integrals is small.  To do so,
we compute the reionization of our cosmological test case assuming
$\taumax =$ 1, 5, 6, and 1000 and using the hybrid treatment for the
periodic depth $\nd$ (see Section~\ref{sec:fedd-lcopt-nd}).  We compare
in Figure~\ref{fig:test_taumax} the resulting errors.

The top panel shows how the median fractional error in the local photon 
number density $\mathcal{J}$ varies with integration time, where we compute 
the fractional errors by comparing against the $\taumax=1000$ test case.  
At early times ($\log(\xhi) > -0.5$) the fractional errors are small 
regardless of the choice of $\taumax$.  This simply reflects the fact 
that, at this epoch, reionization is dominated by scales below our spatial 
resolution so that radiative transfer between cells is subdominant to 
self-ionization of individual cells.  As $\log(\xhi)$ drops below -0.5, 
however, the transport of photons between cells becomes more important and 
the errors in the LC calculation become noticeable.  The errors reach a 
maximum at the point when the individual \hii\ regions begin to overlap 
($\log(\xhi)\sim-1$), and then begin to decline slowly as the universe 
becomes increasingly optically thin.  The slow decline in errors at late 
times owes to the fact that cells can ``see" more sources and the 
Eddington tensors become more nearly isotropic irrespective of 
$\taumax$.  Comparing the error trends for different values of $\taumax$, 
we find that the accuracy errors seem nearly converged even for 
$\taumax = 1$.

The bottom panel shows the distribution of accumulated fractional errors
in local photon number density at the point where the neutral fraction
$\xhi$ has dropped to 7\%.  There is a peak in the error distribution 
near 10-20\%, with significant tails out to low errors and a few regions 
with errors of order unity.  We find no correlation between the magnitude
and fractional error in $\mathcal{J}$; in other words, bright regions
are equally as likely as faint ones to suffer large fractional errors.
As in the top panel, the solution seems generally converged even for
$\taumax=1$ although the errors for $\taumax=5$ and 6 are slightly
lower systematically.  Evidently, choosing $\taumax=1$ would be sufficient
to guarantee a median accuracy better than 10\% at all times; however, in
order to be conservative, we choose $\taumax=6$.

\subsubsection{Optimizing the Periodic Depth}
\label{sec:fedd-lcopt-nd}
\begin{figure}
\setlength{\epsfxsize}{0.5\textwidth}
\centerline{\epsfbox{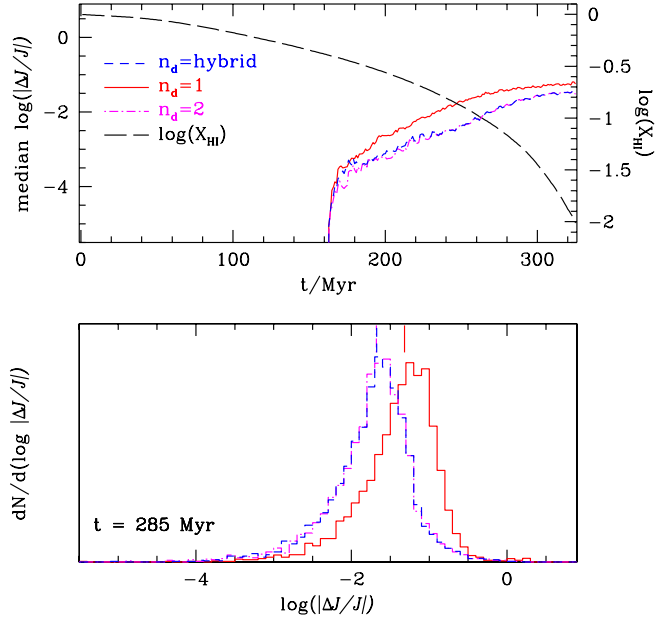}}
\vskip -0.0in
\caption{Convergence-testing the parameter that controls the number of 
simulation volume replicas used to mimic periodic boundaries, $\nd$, 
by comparing fractional errors in the local photon number density $\mathcal{J}$ as 
a function of time.  We compute reionization using $\nd=1$ (red solid), 
$\nd=2$, (magenta dot-dashed), and using a hybrid scheme in which $\nd$ 
switches from 1 to 2 when the volume-averaged neutral hydrogen fraction dips 
below 0.5 (blue short-dashed); the ``converged" answer is obtained by 
assuming $\nd=5$.  The meanings of the various curves are analogous to 
Figure~\ref{fig:test_taumax}.  The hybrid scheme yields median fractional 
accuracy errors of less than 10\% at all times.}
\label{fig:test_nd}
\end{figure}

Hydrodynamical simulations of cosmological volumes assume periodic
boundaries.  We are thus faced with the problem of accounting for periodic
boundaries in our LC calculations.  Unfortunately, an analytic treatment
is impossible as it cannot be determined a priori whether a cell can
``see" a source in a periodic replica of the volume.   The only obvious
treatment is the brute-force approach of mimicking periodic boundaries by
positioning periodic replicas around the simulation volume.  Each source
is then reproduced in each replica, and the LC line integrals must be
computed from each cell in the central volume to each copy of each source.
We use the periodic depth parameter $\nd$ to indicate the depth of the
periodic replicas.  For example, $\nd=1$ corresponds to positioning 26
replica volumes about the simulation.  Clearly, accuracy and computation
time both grow with $\nd$.  Our problem thus reduces to determining the
minimal value of $\nd$ that allows for better than 10\% accuracy at all 
times. 

It is tempting to suppose that, since the universe is optically thick
until the neutral fraction $\xhi$ drops below $\sim10^{-3}$, using
$\nd=1$ should be adequate until then.  However, straightforward testing
indicates that this leads to median errors in photon number density that
approach 10\% even when $\xhi > 5$\%.  In order to determine what value
of $\nd$ leads to converged behavior, we have computed the reionization
of a static density field from $z=9\rightarrow6$ using $\nd=$ 1, 2, 5.
We also introduce a hybrid scheme in which $\nd$ changes from 1 to 2
when the mean neutral fraction drops below 0.5, which, as we will
show, is the best alternative.

In Figure~\ref{fig:test_nd}, we show how the resulting fractional errors
vary with time.  The top panel shows the median fractional error in
local photon number density as a function of time.  We compute the
fractional error by comparing with the $\nd=5$ result, which yields a
very nearly converged solution for low neutral fractions and completely
converged solutions for $\xhi>50$\%.  At early times ($\xhi > 0.6$), the 
universe is so optically thick that few cells are affected by sources in 
replica volumes.  In this regime, using $\nd=1$ leads to negligible errors.  
As $\xhi$ approaches 0.5, however, some of the \hii\ regions grow to 
substantial fractions of the simulation volume and cross its boundaries so 
that the median fractional error begins to rise.  By the time 
$\xhi = 0.01$ ($z\approx 6.3$), the median fractional error exceeds 1\% 
even if $\nd=2$.  Not surprisingly, the errors from $\nd=1$ are larger at 
all timesteps.

The bottom panel shows the distribution of accumulated fractional errors
in local photon number density at the point where the neutral fraction
has dropped to 7\%.  There is a peak in the error distribution near
5--10\%, with significant tails out to low errors and a few regions
with fractional errors of order unity.  The errors in the hybrid scheme
are comparable to the errors for $\nd=2$, with a median error of 3\%.
Noting that the errors generally increase with timestep, we conclude
that the hybrid scheme leads to median errors that are always $\leq 10\%$.

The hybrid $\nd$ scheme speeds up our code significantly.  We have
verified through direct testing that the LC computation time varies with
the periodic depth roughly as $(2\nd + 1)^3$ (that is, proportional
to the total number of volumes), regardless of both the spatial
resolution and the ionization state of the universe.  It follows that,
for a calculation in which $\xhi$ drops below 50\% after roughly half
of the total integration time has elapsed, using hybrid $\nd$ rather
than using $\nd=2$ reduces the computation time by up to $\approx$40\%
without affecting the accuracy.  Hence this is the preferred option.

\subsubsection{Optimizing the Eddington Tensor Update Criterion}
\label{sec:fedd-lcopt-fc}

\begin{figure}
\setlength{\epsfxsize}{0.5\textwidth}
\centerline{\epsfbox{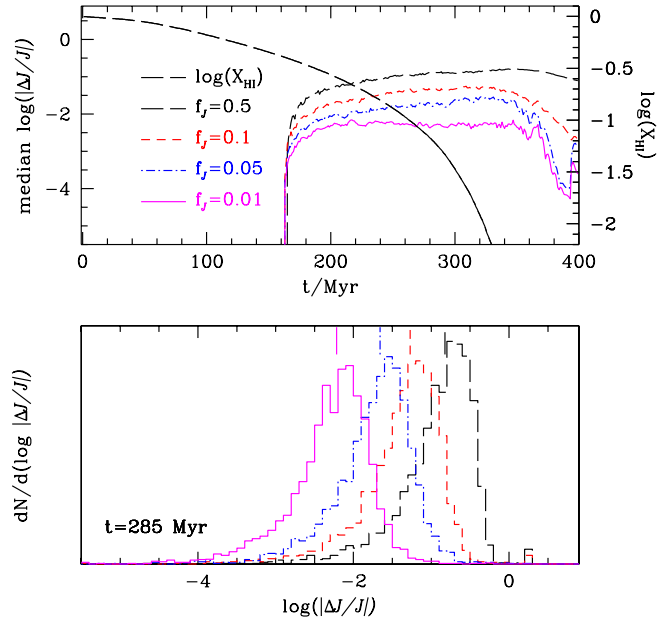}}
\vskip -0.0in
\caption{Convergence-testing the parameter that controls the frequency of
updates to the Eddington tensor, $\fJ$, by comparing fractional errors in 
the local photon number density $\mathcal{J}$ as a function of time.  We 
compute reionization using $\fJ=0.5$ (black long-dashed), $\fJ=0.1$ (red
short-dashed), $\fJ=0.05$ (blue dot-dashed) and $\fJ=0.01$ (magenta 
solid); the ``converged" answer is obtained with $\fJ=0$.  The meanings 
of the various curves are analogous to Figure~\ref{fig:test_taumax}.  
Choosing $\fJ=0.05$ yields median errors better than 10\% at all times.}
\label{fig:test_fJ}
\end{figure}

Our technique involves periodically updating the Eddington tensor field 
in order to maintain consistency with the time-dependent
integration.  Naturally, more frequent Eddington tensor updates lead to
a more accurate solution; in fact,~\citet{aue70} recommend iterating
to convergence between the radiation and Eddington tensor fields.
Unfortunately, this approach would be prohibitively time-consuming for
our problem.  Instead, we opt to update the Eddington tensor in a
given computational cell only when the radiation field has changed 
significantly in that cell; this technique has been shown to be an 
excellent approximation in other contexts~\citep[for example,][]{hub07}.   
In particular, after each timestep, we re-compute the Eddington tensor 
in those cells where the photon number density has undergone a 
fractional change greater than $\fJ$ in at least one frequency bin 
since the last update to its Eddington tensors.  In order to determine 
how the resulting errors vary with $\fJ$, we compute the reionization 
of a static density field from $z=9\rightarrow6$ using $\fJ=$ 0, 0.01, 
0.05, 0.1, and 0.5.

In Figure~\ref{fig:test_fJ} we show how the resulting fractional errors
vary with time.  This figure demonstrates the power and flexibility of 
the moment method: even if we only compute the full angular dependence 
of the radiation field when it has changed by more than a factor of two
($\fJ=0.5$), the typical local errors are better than 20\% at all times.  
Most importantly, because setting $\fJ>0$ results in less-frequent 
updates to the Eddington tensor field, it speeds up the computation
considerably.  Through direct testing, we find that the computation 
time $t$ for our reionization calculation varies with $\fJ$ as 
$t \propto \fJ^{-0.5}$.

We now examine the top panel of Figure~\ref{fig:test_fJ} in more detail.  
At early times ($\xhi > 0.5$), errors are small because reionization is 
dominated by the self-ionization of small, overdense regions rather than 
by radiation transport.  Eddington tensor updates are frequent in overdense 
regions owing to the rapidly-evolving $\mathcal{J}$, but they are also fast 
because, in an optically thick universe, the $\taumax$ parameter insures 
that most of the LC line integrals terminate well before they arrive at the 
source.  After $\xhi$ drops below 0.5, the errors begin to grow.  However, 
rather than growing without bound, they level off at a characteristic value 
that in turn grows with $\fJ$.  This is the regime in which setting $\fJ>0$
yields the biggest savings in computation time because, on average, the 
LC line integrals traverse more cells before reaching $\taumax$.  On the 
other hand, the radiation field evolves more slowly because the 
rapidly-evolving overdense regions are already largely reionized.  Hence 
setting $\fJ>0$ corresponds to culling the most expensive Eddington 
tensor updates aggressively while preserving the overall accuracy.

The bottom panel of Figure~\ref{fig:test_fJ} indicates that the distribution
of errors shifts to smaller errors with decreasing $\fJ$, with values of 
$\fJ$ less than $0.1$ leading to median errors $\leq 10$\%.  In practice, 
errors may be slightly larger as the median error at a given timestep and 
$\fJ$ increases slightly with increasing spatial resolution.  Moreover, 
Figure~\ref{fig:test_fJ} does not address errors in the topology of the 
ionization field, which may vary differently with $\fJ$ 
(although~\citealt{mcq07} argue that the topology of reionization is 
dominated by the emissivity field and $\xhi$ and does not vary strongly 
with other parameters).  The appropriate choice of $\fJ$, therefore, depends 
on the problem.  However, for reference, we note that choosing $\fJ = 0.05$ 
generally leads to median errors that are better than 10\% while speeding 
up the calculation by roughly a factor of three.

\subsubsection{Full Error Budget}
\label{sec:fedd-lcopt-full-err}

\begin{figure}
\setlength{\epsfxsize}{0.5\textwidth}
\centerline{\epsfbox{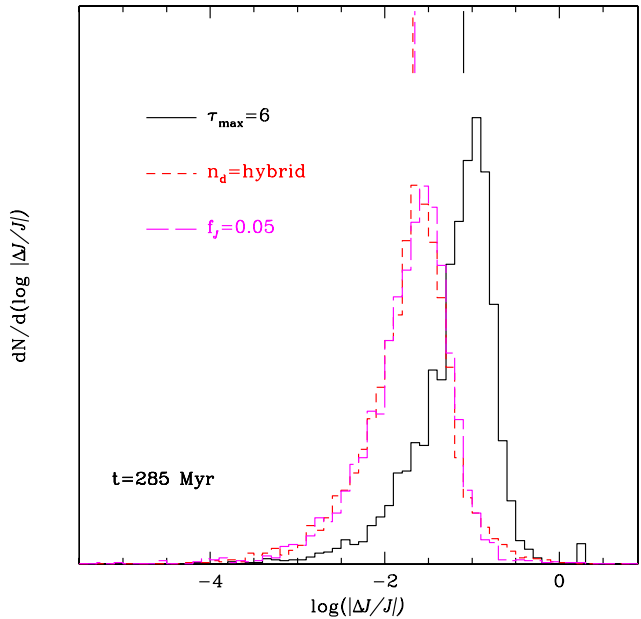}}
\vskip -0.0in
\caption{The distributions of accuracy error at $t=285$ for our fiducial
choice of numerical parameters ($\taumax$, $\nd$, $\fJ$) = 
(6, ``hybrid", 0.05).  For this set of parameters, the error is dominated 
by the error owing to our choice of $\taumax$.  Adding the median errors
in quadrature, we expect a typical error of 9\% in $\mathcal{J}$.}
\label{fig:error_budget}
\end{figure}

In Figure~\ref{fig:error_budget}, we summarize the results of these 
convergence tests by comparing the distributions of errors in accuracy
at $t=285$ that result from our fiducial choice of numerical parameters.
The median errors for $\taumax=6$, hybrid $\nd$, and $\fJ=0.05$ are
8.0, 2.1, and 2.2\%, respectively.  Hence our fiducial choice of 
parameters leads to a typical numerical accuracy of 10\% in $\mathcal{J}$.

\subsection{Computational Scaling}
\label{sec:fedd-lc-scaling}
\begin{figure}
\setlength{\epsfxsize}{0.5\textwidth}
\centerline{\epsfbox{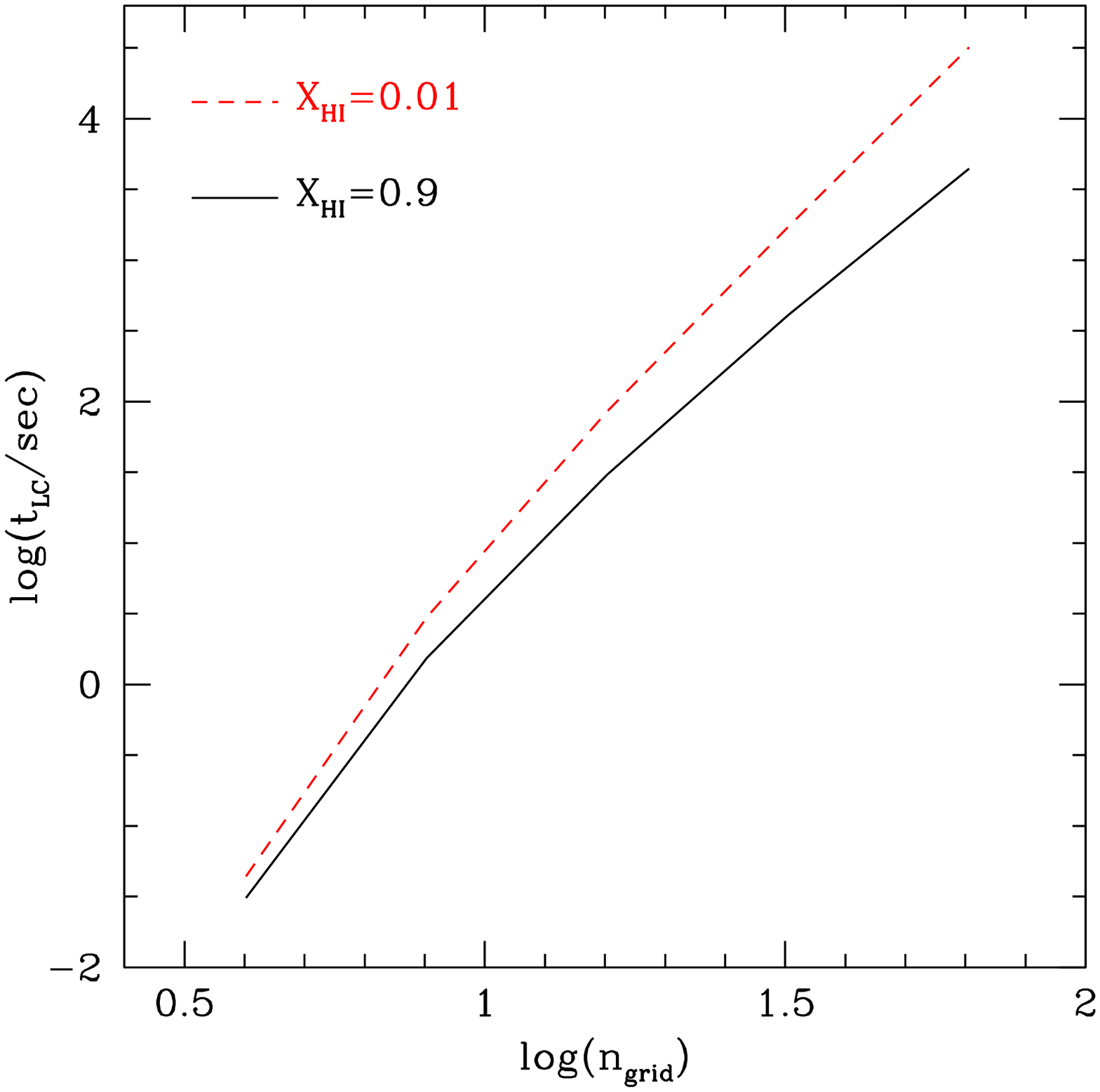}}
\vskip -0.0in
\caption{LC computation time versus number of grid cells to a side in 
the case of a mostly neutral (solid black) and highly ionized (dashed red) 
universe.  The scaling flattens from $\ngrid^{5.6\mbox{--}6.1}$ at coarse 
resolution to $\ngrid^{3.4\mbox{--}4.2}$ at high resolution, depending
on the neutral fraction $\xhi$.}
\label{fig:test_lc_scaling}
\end{figure}
The computation time for LC, $\tlc$, is proportional to the 
number of cells $\ngrid^3$, the number of cells containing sources $n_S$, 
and the average length of a line integral in cells $n_l$:
\begin{equation*}
\tlc \propto \ngrid^3 n_S n_l
\end{equation*}
Both $n_l$ and $n_S$ generally vary with spatial resolution as well as 
with the mean opacity.  In the limit of an optically thin volume 
($n_l \sim \ngrid$) with nonzero emissivity everywhere ($n_S = \ngrid^3$),
$\tlc \propto \ngrid^7$, while in the limit of an optically thick 
medium ($n_l$ constant) and highly clustered sources ($n_S$ constant) 
the scaling flattens to $\tlc \propto \ngrid^3$.  Because $\tlc$ 
potentially scales quite unfavorably with $\ngrid$, we expect that it 
will ultimately dominate our spatial resolution limit.  Hence, it is 
of interest to determine where the scaling falls for our problem.  To 
do so, we gridded the same cosmological snapshot used in 
Section~\ref{sec:fedd-lcopt} onto grids of increasing spatial resolution 
assuming two different uniform ionized fractions, $\xhi=0.9$ and 
$\xhi=0.01$.  We then ran our LC code on these fields and used the gnu 
gprof utility to measure the computation time.  We repeated the 
low-resolution tests up to ten times and averaged the results.  For 
consistency, we set the periodic depth parameter to $\nd=1$ in all 
cases.  We performed the calculations on a single 2-GHz AMD Athlon 
processor.  In Figure~\ref{fig:test_lc_scaling} we compare the 
resulting scalings.  

Looking at the nearly-neutral case ($\xhi=0.9$) first, we see that 
at coarse resolution our code scales as $\ngrid^{5.6}$,
indicating that $n_S$ is increasing rapidly with increasing 
resolution.  At higher resolutions the scaling flattens to 
$\ngrid^{3.4}$, indicating both that $n_S$ is varying only very slowly
with resolution owing to source clustering and that $n_l$ is varying 
more slowly than $\ngrid$ because most LC line integrals are terminated 
at the first cell owing to the high opacity.

Turning to the nearly-ionized case ($\xhi=0.01$), we find that the 
computation times are uniformly higher and the scaling is slightly
steeper than in the nearly-neutral case, flattening from 
$\ngrid^{6.1}$ at coarse resolution to $\ngrid^{4.2}$ at high resolution.  
Because $n_S$ is the same in both cases, these differences owe entirely 
to changes in $n_l$.  First, in a more optically thin volume $n_l$ is 
greater overall because, on average, line integrals traverse more cells 
before reaching $\taumax$.  Second, $n_l$ scales more nearly as $\ngrid$ 
because more of the line integrals proceed past the first cell even at 
low spatial resolution.

In summary, we find that, for the problem of cosmological reionization, 
$\tlc \propto \ngrid^{3.4\mbox{--}6.1}$, depending on $\xhi$ and 
$\ngrid$.  The flat scaling at high $\xhi$ and high spatial resolution is 
encouraging, and indicates that LC should lend itself well to studying 
galaxy evolution well before the epoch of overlap ($z=$6--7).  On the 
other hand, the generally longer computation times and steeper scaling at 
low $\xhi$ indicate that our code will slow down considerably as the mean 
neutral fraction drops below $0.01$.  In future work we plan to study how 
to transition to the much faster optically thin approximation at this 
epoch without introducing significant inaccuracies.

\section{Solving for the Non-Equilibrium Ionization States} \label{sec:nlte}
\begin{figure}
\setlength{\epsfxsize}{0.5\textwidth}
\centerline{\epsfbox{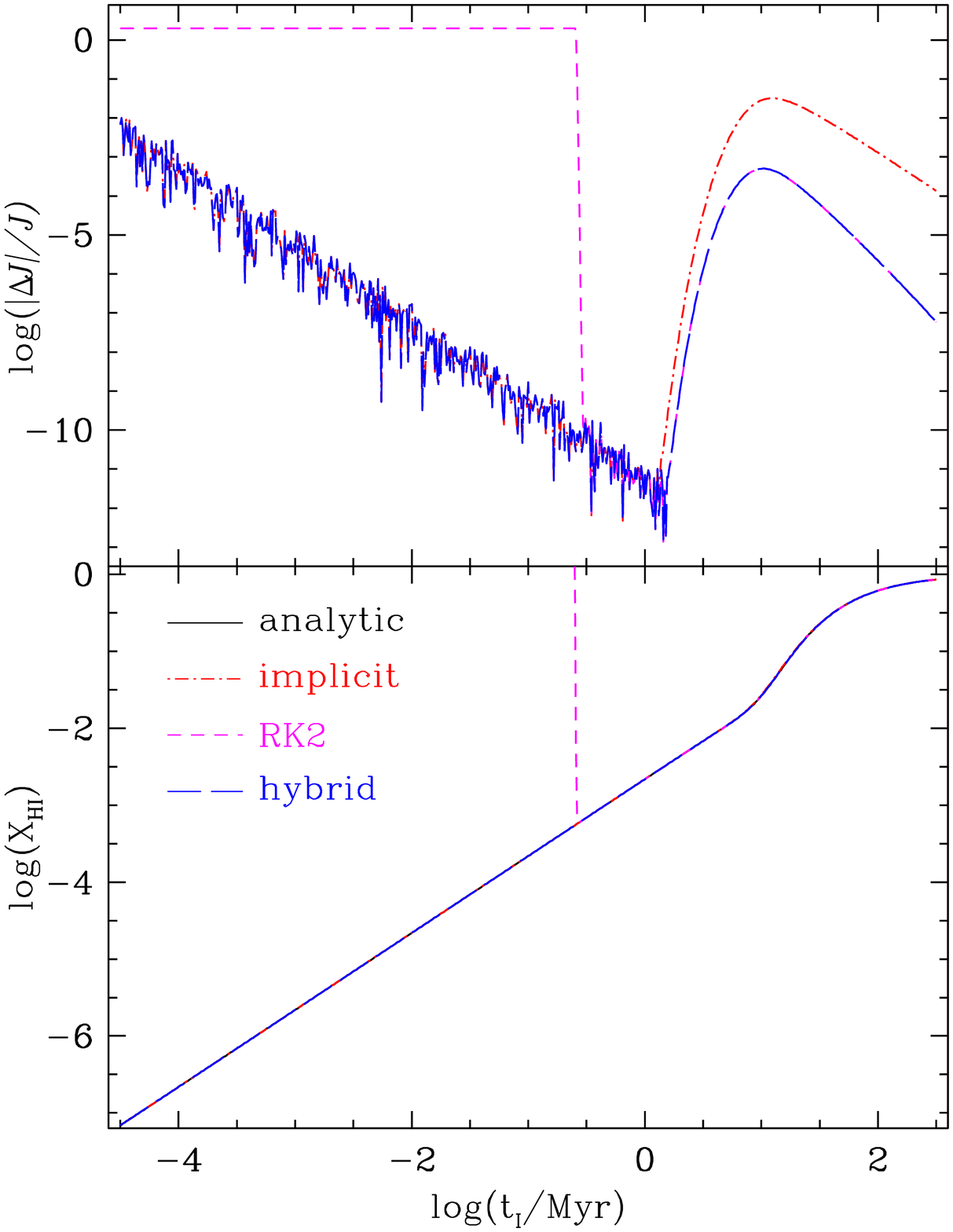}}
\vskip -0.0in
\caption{(top) Fractional error in final ionization state as a function
of ionization timescale for three different nonequilibrium integration
techniques.  (bottom) Final neutral fraction for the same tests.  For 
short ionization timescales the error is dominated by roundoff error 
while for long ionization timescales the accumulated error is 
dominated by timestep truncation error.  A hybrid scheme using 
implicit differencing for short timescales and RK2 for long 
timescales is accurate to better than 1\% in all regimes of 
interest.
}
\label{fig:test_nlte}
\end{figure}
In order to compute cosmological reionization, we must integrate the 
nonequilibrium equations for ionization and recombination of hydrogen
and helium.  A thorough discussion of the relevant chemical processes
including analytic fits for the cross sections and reaction rates is 
provided by~\citet{ann97} and~\citet{abe97}.  Here, we focus on deriving 
a technique for integrating these equations.  We neglect H$^-$ and H$_2$ 
because we do not anticipate being able to resolve the mass scales at 
which these species are expected to dominate 
($<10^{8}\msun$; see, e.g.,~\citealt{cou86}).  However, it would 
be trivial to extend our technique to account for these species as well.

Following~\citet{ann97}, we write the equation that governs the 
abundance of species $i$ schematically as
\begin{equation}
\label{eqn:nlte}
\frac{\partial n_i}{\partial t} = C_i(T,n_j) - D_i(T,n_j) n_i,
\end{equation}
where $C_i$ and $D_i$ respectively represent source and destruction 
terms summed over all species $j$ that can convert to or result from
species $i$.  Equation~\ref{eqn:nlte} can be modified to hold in an 
expanding universe by redefining $n_i$ as the comoving number density 
and normalizing the reaction rate coefficients by $a^3$.

As is well known, Equation~\ref{eqn:nlte} can be quite stiff in the 
traditional sense; that is, its individual terms can evolve on very 
different timescales.  This property has led many authors to adopt 
unconditionally stable integration 
techniques~\citep[see, e.g.,][]{ann97,mel06}.  Unfortunately,
stable techniques are not necessarily accurate.  Moreover, within
cosmological density fields the timescales of the terms in 
Equation~\ref{eqn:nlte} can vary rapidly with position.  For example, 
we have found that the timescales encountered during a typical 
reionization calculation at our expected spatial resolution can vary 
between $10^{-4}$--$10^2$ Myr at a given time.  We have explored the 
stability and accuracy of a number of integration techniques when 
applied to a range of gas densities and ionization timescales in order 
to derive an optimal technique.  Here we discuss two techniques, a 
fully implicit (FI) backwards differencing formula and second-order 
Runge-Kutta (RK2).

Backwards differencing Equation~\ref{eqn:nlte} results in the following
difference equation for the updated density $n_i^{n+1}$ in terms of the 
previous density $n_i^n$ and the updated densities of the other species
$n_j^{n+1}$:
\begin{equation}
\label{eqn:bdf}
\frac{n_i^{n+1}-n_i^{n}}{\Delta t} = C_i(T,n_j^{n+1})-D_i(T,n_j^{n+1})n_i^{n+1},
\end{equation}
where all densities on the right hand side are evaluated at the updated
time.  This fully implicit technique is accurate to first order in time 
and is stiffly stable~\citep{gea71}.  We solve the resulting set of 
algebraic equations using Newton-Raphson iteration.  For 1 Myr timesteps, 
we have found that the iterative solution converges to within a tolerance 
of $10^{-4}$ in fewer than 4 iterations, 
hence it is also reasonably efficient.  However, evaluating and inverting 
the Jacobian is time-consuming.  Moreover, in regimes where the shortest 
timescale is comparable to the timestep, its accuracy suffers because the 
rate coefficients at the end of the timestep are not a good approximation 
for their timestep-averaged values.

Second order explicit Runge-Kutta methods are accurate to second order 
in time and can be computed rapidly, but they become unstable if the 
timestep is comparable to or longer than the shortest relevant timescale.  
We have implemented the standard form of RK2~\citep[e.g.,][]{nr}.
In addition, we have implemented a hybrid technique that combines the 
accuracy of RK2 around long timescales with the stability of implicit 
methods at short timescales.  To do so, the hybrid technique simply 
checks all relevant timescales before each timestep and uses RK2 
whenever the timestep is shorter than the shortest relevant timescale.
 
We test these techniques by solving Equation~\ref{eqn:nlte} for a 
single zone composed of pure hydrogen.  The zone has a uniform, constant 
total number density and is initially entirely neutral.  The ionizing 
radiation field is time-independent, and we refer to its intensity by 
its associated ionization timescale $t_I$.  The relevant processes are 
radiative ionization, collisional ionization owing to collisions with 
electrons, and radiative recombination assuming case-B recombination 
rates.  We evolve the ionization state using 0.5-Myr timesteps for 
50 Myr and then determine the accumulated fractional error in the 
numerical result by comparing to the analytical solution.  The total 
number density is $1.66\times10^{-4}$ cm$^{-3}$, roughly the 
cosmological mean density at $z=9$, and we ignore Hubble expansion 
for simplicity.

In Figure~\ref{fig:test_nlte} we compare the results of performing the test
integration with our three schemes versus the analytical solution.  In
the top panel, we find that the accumulated fractional error varies 
nontrivially with $t_I$.  For small $t_I$ (i.e., intense ionizing 
backgrounds) the solution reaches ionization equilibrium well before
the integration ends.  In this ``equilibrium regime" the error grows with 
decreasing $t_I$.  We have found that the accuracy here cannot be further 
improved by integrating with smaller timesteps or switching to a 
higher-order finite-differencing scheme in time (for example, note that 
the second-order accurate RK2 scheme yields the same magnitude of error 
as the first-order accurate FI scheme for $\log(t_I) = -0.5$--1).  The 
``error floor" owes directly to roundoff error: if the machine 
accuracy is $\epsilon_{\mathrm{m}}$ and if the solution is close enough 
to equilibrium that the change in the next timestep is small compared to 
the current solution, $|\dot{n}(n_i)|\Delta t < \epsilon_{\mathrm{m}}n_i$,
then the error cannot decay.

For large enough $t_I$ (weak ionizing backgrounds), the solution does 
not reach ionization equilibrium and roundoff error becomes subdominant.  
In this ``nonequilibrium regime" the error is dominated by the usual 
truncation error in the finite-differencing scheme with the result that 
using shorter timesteps or switching to a higher-order finite differencing 
scheme improves the accuracy, as can immediately be seen from
Figure~\ref{fig:test_nlte}.  We repeated this test integration with 
a range of densities, timesteps, and integration times.  In the 
nonequilibrium regime, we find that RK2 consistently yields accuracies 
better than 1\% while FI often yields fractional errors of order unity.

To evaluate how much computation time we save through our hybrid technique,
we tracked the number of calls to each of the two ionization routines
throughout a test calculation of reionization similar to that in
Section~\ref{sec:test_cosmo} but with only $8^3$ computational cells.  
Before reionization ($\xhi >$50\% ) RK2 is chosen 91\% of the time, while 
after reionization it is chosen 38\% of the time.  Overall, RK2 is called 
52\% of the time.  Noting that RK2 is $\approx 4$ times faster than FI, 
our use of a hybrid technique rather than relying only on FI roughly 
halves the time required for evolving the ionization state.  In other 
words, it simultaneously improves both the accuracy \emph{and} the 
efficiency of our technique.

The bottom panel can be used to determine whether our tests span the
domain of physical conditions that arise in cosmological calculations.  
At our anticipated spatial resolution ($>50\hkpc$), $t_I$ ranges between 
$10^{-4}$ and $10^2$ Myr while the final (that is, post-recombination) 
hydrogen neutral fraction ranges between $10^{-7}$ and 1.  Our test 
calculations clearly span this domain, hence we conclude that our hybrid 
integration technique stably and efficiently yields fractional errors of
$\leq 1\%$ throughout our problem's domain.

\section{Putting it All Together} \label{sec:full_alg}

\begin{figure}
\setlength{\epsfxsize}{0.5\textwidth}
\centerline{\epsfbox{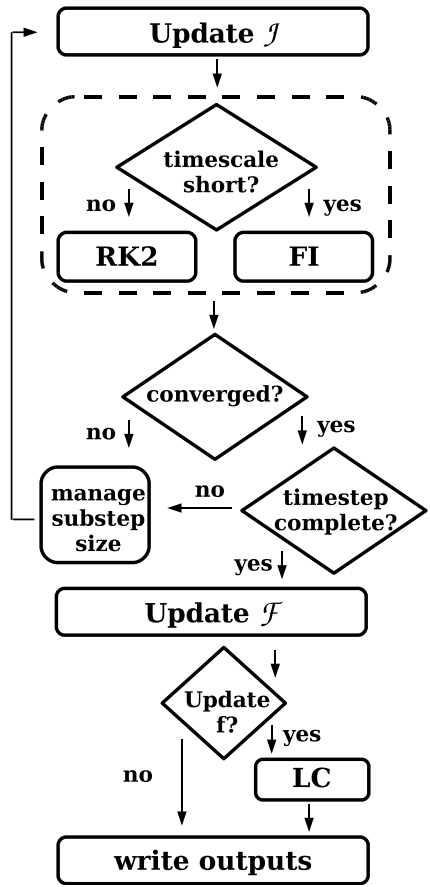}}
\vskip -0.0in
\caption{A single timestep in our code.}
\label{fig:flowchart}
\end{figure}

Having discussed our techniques for updating the Eddington tensor,
radiation, and ionization fields, we now turn to our method for 
combining these ingredients into a single code.  
Figure~\ref{fig:flowchart} illustrates our algorithm for computing 
a single timestep.  At the beginning, we solve self-consistently 
for the updated photon number densities $\mathcal{J}^{n+1}$ and
ionization states $n^{n+1}$ in terms of the previous values
($\mathcal{J}^n$, $\mathcal{F}^n$, $n^n$) through iteration.  
Schematically, we loop through the following calculations until the 
solutions have converged:
\begin{eqnarray*}
\mathcal{J}^{n+1} & = & \mathcal{J}^{n+1}(\mathcal{J}^n,n^{n+1},\mathcal{F}^n)  \\
n^{n+1} & = & n^{n+1}(\mathcal{J}^{n+1},n^n)
\end{eqnarray*}
During the first iteration of each timestep, we use the values from 
the previous timestep as the initial guess for the updated values.

Because this scheme does not converge for general initial conditions 
and timestep $\Delta t$, we must include a treatment for reducing the 
timestep whenever necessary.  We have implemented an adaptive stepsize 
scheme that is designed to collapse the timestep rapidly near difficult 
spots but then slowly accelerate as the integration grows smoother.  
In particular:
\begin{enumerate}
\item If the fractional error is less than $10^{-4}$, then we consider 
the solution to have converged.  We halt the iteration and advance
the timestep.
\item If the fractional error does not reach $10^{-4}$ in 15 
iterations, we divide $\Delta t$ by 4 and restart the iteration.
\item If 4 consecutive substepped iterations converge, then we
multiply $\Delta t$ by 2.
\item If the substepped timestep becomes smaller than $10^{-5}$ of the 
original timestep, then we consider the computation to have diverged.
In this case, we terminate the integration.
\end{enumerate}
After updating the radiation and ionization fields, we compute the
updated fluxes $\mathcal{F}^{n+1}$.  We then compute the fractional
change in each cell's $\mathcal{J}$ since the last update to its Eddington
tensor and update the Eddington tensors wherever the fractional change 
exceeds $\fJ$.  This marks the end of a single timestep.

\section{Tests} \label{sec:tests}
\subsection{Str\"omgren Spheres}
\begin{figure*}
\centerline{
\begin{tabular}{rl}
\setlength{\epsfxsize}{0.5\textwidth}
\centerline{\epsfbox{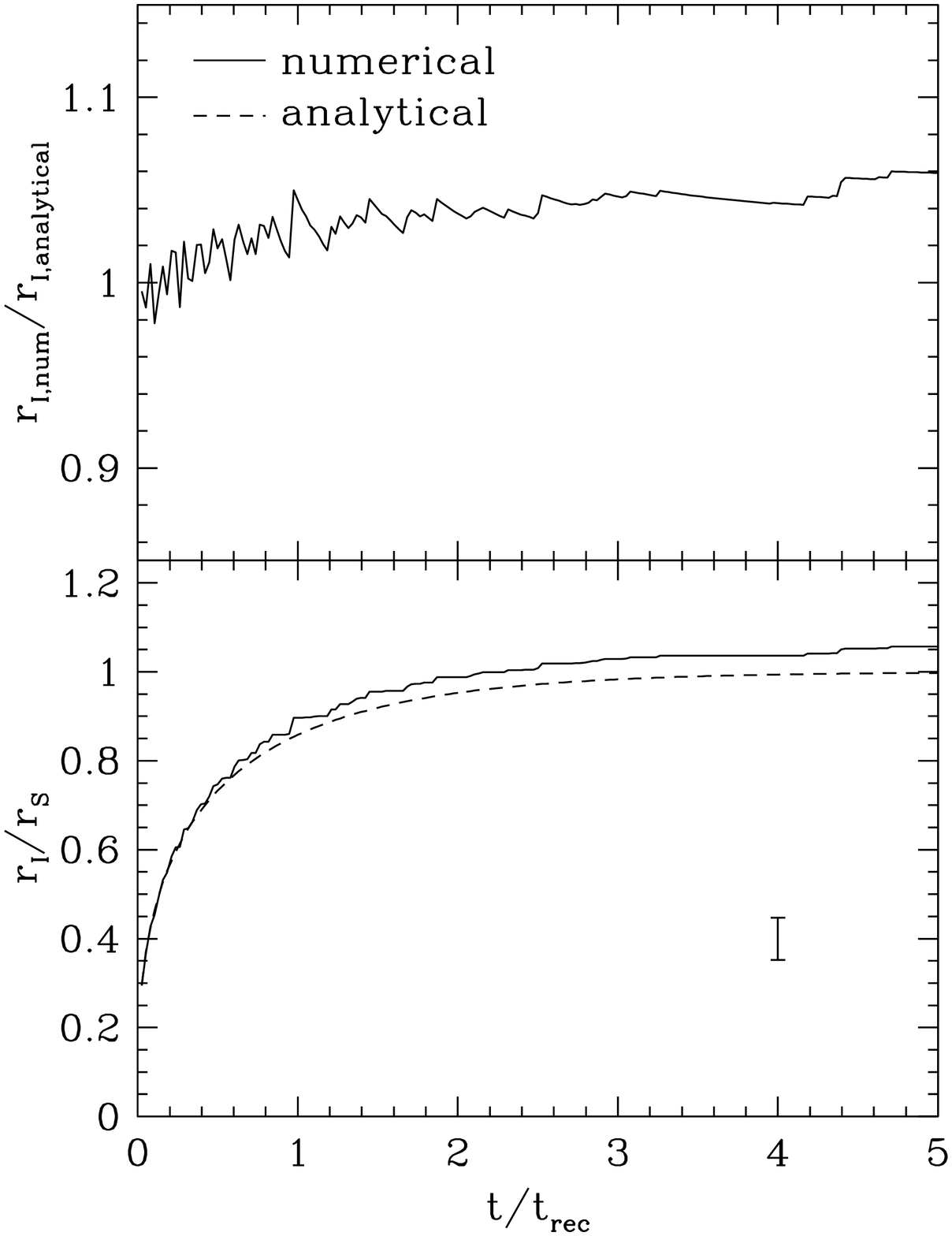}}
&
\hskip -0.5\textwidth
\setlength{\epsfxsize}{0.5\textwidth}
\centerline{\epsfbox{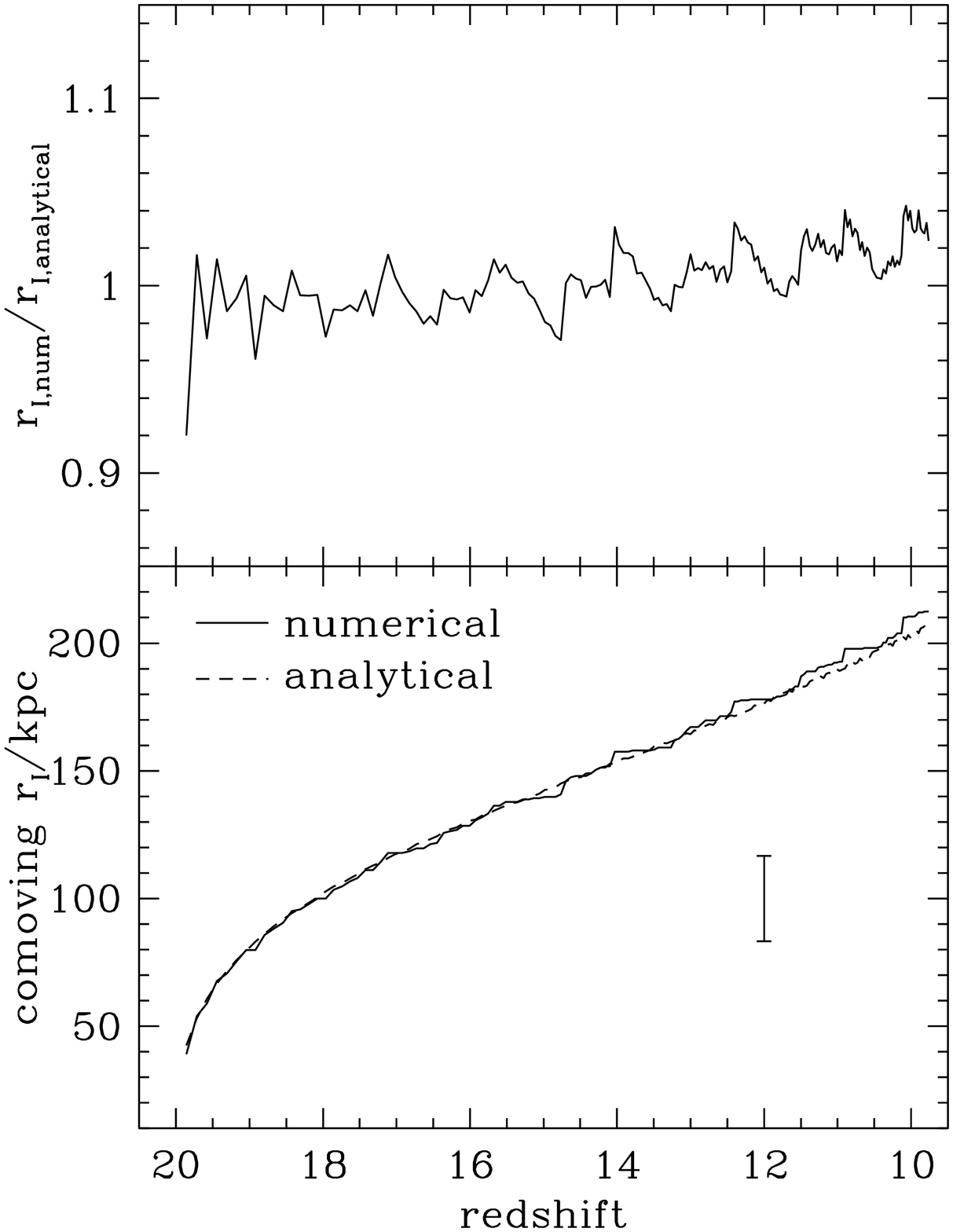}}
\end{tabular}
}
\caption{Test of classical (left) and cosmological (right) Str\"omgren 
spheres.  In the bottom panels we show the numerical (solid) and 
analytical (dashed) solutions as a function of time.  The error bars
in the bottom panel are included for reference and span twice the 
width of a computational cell.  In the top panels we show the ratio 
of the numerical to the analytical solutions.  In both test cases, 
the numerical and analytical solutions agree to within one cell width 
at all times.}
\label{fig:stromgren}
\end{figure*}

We now demonstrate that our code accurately computes the growth of
\hii\ regions in both static and expanding media.  For the static
case, we locate a single O star of monochromatic ionizing luminosity 
$5\times10^{48}$ s$^{-1}$ in a homogeneous medium of pure hydrogen with 
total density $10^{-3}$ cm$^{-3}$, temperature $10^4$ K, and initial 
ionized fraction 0.0012 (Test 1 from~\citealt{ili06}).  The simulation
volume has a side length of 10.5 kpc and is divided into $48^3$ cells.  
We evolve the nonequilibrium radiation and ionization fields for 500 Myr.  
In the bottom-left panel of Figure~\ref{fig:stromgren}, we compare how
the radius of the resulting ionization front grows with time in our code 
(solid) versus the exact analytical solution~\citep[dashed;][]{ili06}.
In the numerical case, we define the radius of the ionized region as the 
radius at which the neutral hydrogen fraction drops to 50\%.  

There is a slight tendency for the numerical radius to exceed the 
analytical one owing to the diffusive nature of the moment method (recall 
that the divergence in Equation~\ref{eqn:rt-moments-tstep-E} is computed 
from a 19-cell finite difference pattern).  In order to put this effect
into context, we include an error bar that spans twice the width of a 
single computational cell; this is our true spatial resolution.  Comparing 
the error bar with the gap reveals that our method is accurate to within 
the uncertainty introduced by the finite spatial resolution.  In the top 
panel we show the ratio of the numerical to the analytical solutions as a 
function of time.  At all times, they agree to within 5\%.  Noting that 
the majority of the codes that are compared in~\citet{ili06} also yield 
ionization front radii that exceed the analytical solution by 1--5\%
(Figure 7 of~\citealt{ili06}), we conclude that our method's accuracy is 
comparable to those codes.

For the expanding case, we consider a protogalaxy that ``turns on"
at $z=20$ with an ionizing luminosity $5\times10^{49}$ s$^{-1}$ (this 
is roughly what is expected for $10^4 \msun$ of young Population II 
stars assuming an ionizing escape fraction of 5\%).  The protogalaxy
lives in a homogeneous medium of pure hydrogen with comoving density
$1.66\times10^{-7}$ cm$^{-3}$ and temperature $10^4$~K.  We evolve 
the test from $z=20\rightarrow10$ in an Einstein-de Sitter universe 
with $h=0.7$ using $48^3$ computational cells.  In the bottom right 
panel of Figure~\ref{fig:stromgren}, we compare the comoving radius of 
the ionized region as a function of time in our numerical model (solid) 
against the analytical solution of~\citet{sha87} (dashed).  As before, 
the numerical solution tracks the analytical one to within the size of 
a grid cell at all times.  In the top right panel we show the ratio of
the numerical to the analytical solutions as a function of time.  Once
again, our numerical solution is accurate to within 5\% at all times.

Combining the results of these tests, we conclude that our code conserves
photons, accurately determines the nonequilibrium ionization states, and 
accounts for the relevant cosmological terms.

\subsection{Shadowing}
\begin{figure*}
\centerline{
\setlength{\epsfxsize}{0.9\textwidth}
\centerline{\epsfbox{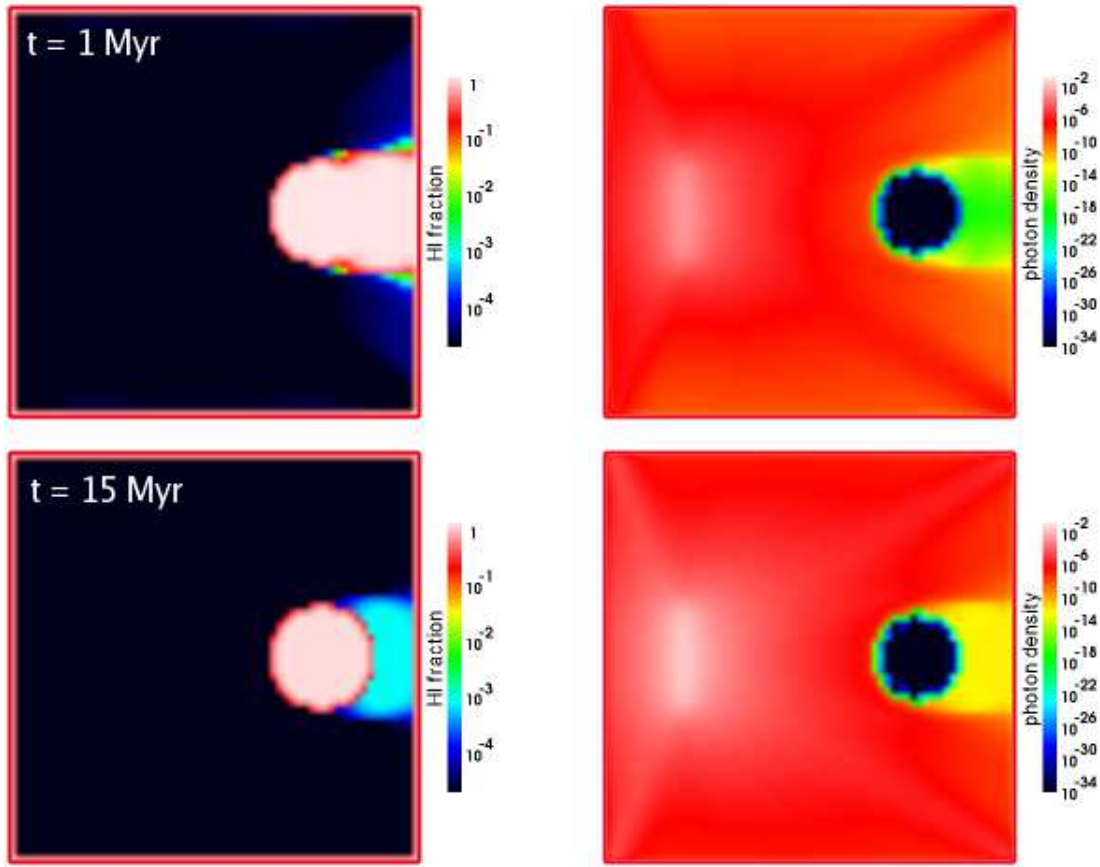}}
}
\caption{Test of ionization front trapping: a luminous disk irradiates a 
dense clump of the same radius in a diffuse medium.  The left and right 
panels show the neutral hydrogen fraction and photon number density,
respectively, in a plane cut through the middle of the simulation volume 
at $t=1$ (top) and 15 (bottom) Myr.  The brightest areas give the position 
of the source, and the darkest areas give the position of the barrier.  
The ``spokes" radiating from the source to the corners in the right panels 
are a numerical artifact of the opaque boundary with which we surround 
the simulation box in order to suppress periodic effects.  Our moment 
method results in incomplete shadowing owing to numerical diffusion.
These Figures were produced using {\sc ifrit}.}
\label{fig:shadow}
\end{figure*}

We now discuss how well our code is able to produce shadows.  Our goal is
to run a test case whose results can be compared to Test 3 of~\citet{ili06}.
However, this test involves irradiating a dense clump with a plane-parallel 
wavefront, a situation that is difficult to impose in our method.
Instead, we consider the case of a dense clump of cold hydrogen 
irradiated by a bright disk located sufficiently far away that the flux 
from the disk at the clump is approximately plane-parallel.  We calibrate
the disk's (isotropic) emissivity to the simulation resolution so that, if 
it were an infinite plane, the flux would be $2\times10^6$ s$^{-1}$ 
cm$^{-2}$.  The radius of the clump and the disk is 0.8 kpc, they are 
separated by 3.75 kpc, and the plane of the disk is oriented perpendicular 
to the line connecting the disk and the clump.  The ambient hydrogen number 
density and temperature are $2\times10^{-4}$ cm$^{-3}$ and 8000 K, while 
inside the clump they are 0.04 cm$^{-3}$ and 40 K.  The box size is 6.6 kpc 
and contains $64^3$ computational cells.  We evolve the simulation for 15 
Myr.

Figure~\ref{fig:shadow} shows the neutral hydrogen fraction and photon
number density in the simulation mid-plane after 1 Myr (top panels) and 15
Myrs (bottom panels).  In these figures the dense clump is on the right
and the source disk, seen edge on, is on the left.  Looking at the top
panels first, we see that after 1 Myr the region behind the clump remains
largely neutral although the rest of the volume is completely ionized,
suggesting that our technique shadows well, with the photon density at
this time 4--5 orders of magnitude fainter in the shadowed region than
in the unshadowed one.  The small amount of diffusion results from the
fact that our LC module sets the Eddington tensors in the shadowed
region to the isotropic case because no sources are visible there.
Consequently, photons are free to diffuse into the shadowed region,
as is expected in any moment formalism.  After 15 Myr, the ionizing
field behind the clump has strengthened to a photon number density of
roughly 1\% of the value in the unshadowed region, driving the neutral
hydrogen fraction down to $10^{-3}$.  At this point, the volume is in
equilibrium; in fact, the radiation and ionization fields do not change
appreciably between 5 and 15 Myr.

Comparing Figure~\ref{fig:shadow} to Figures 22 and 24 of~\citet{ili06},
we find that our technique shadows as well as ray-tracing and Monte 
Carlo codes at $t=1$ Myr.  However, at $t=15$ Myr our code performs
more poorly owing to the diffusion of photons into the shadowed region.  
Figure~\ref{fig:shadow} can also be compared with~\citet{hay03}, who 
introduced a moment method that is similar to ours.  They demonstrated 
that incorporating Eddington factors from a time-dependent short 
characteristics integration results in incomplete shadowing (see their 
Figures 6--9).  They also found that increasing the spatial resolution 
did not resolve the problem.  Instead, they concluded that the 
incomplete shadowing results from the fourth term in 
Equation~\ref{eqn:rt-moments-tstep-E}, which involves the Laplacian
of the product $\mathbf{f}\mathcal{J}$.  Evaluating this term couples
nonadjacent computational cells and gives rise to numerical diffusion
in space.  This might be expected to present even more of a problem in
our technique given that we find it necessary to smooth our Eddington
tensors whereas they did not.  Despite this, Figure~\ref{fig:shadow}
demonstrates that our technique can shadow quite effectively.

\subsection{Cosmological Volume} \label{sec:test_cosmo}
\begin{figure*}
\centerline{
\setlength{\epsfxsize}{0.9\textwidth}
\centerline{\epsfbox{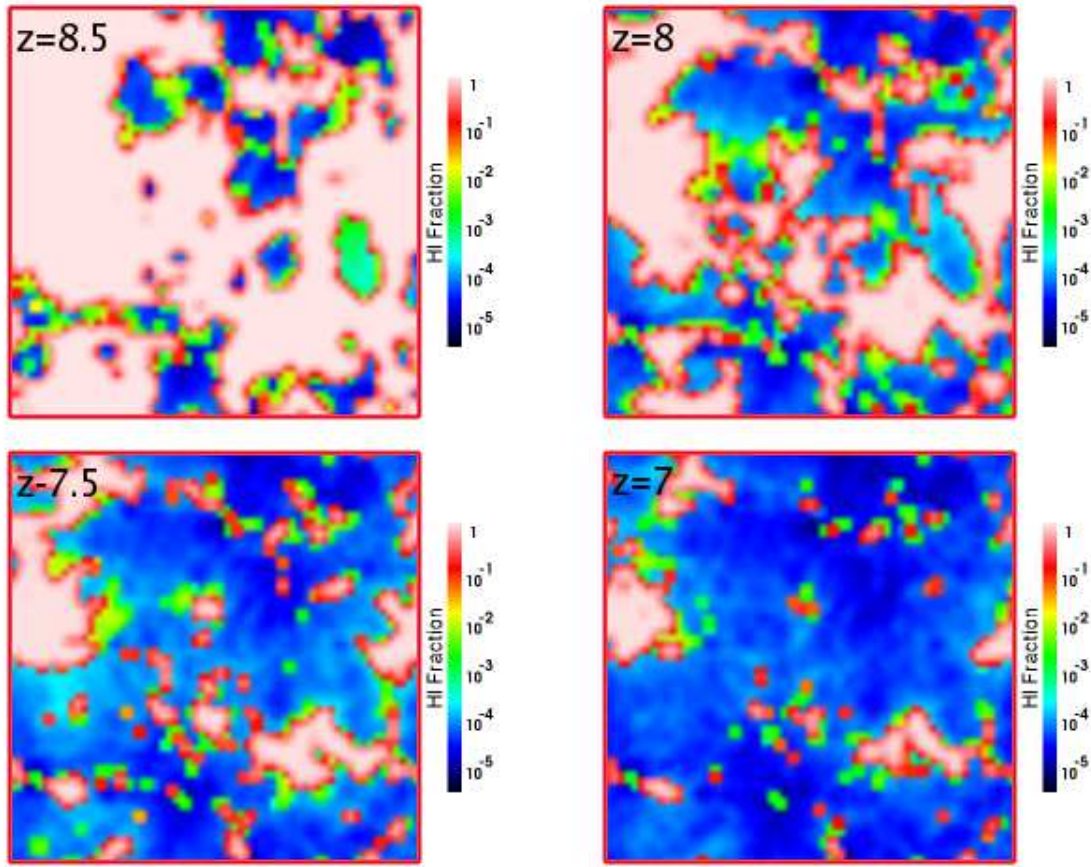}}
}
\caption{Two-dimensional slices through a reionization computation with 
a static emissivity field at four representative redshifts.  The neutral 
hydrogen fractions are color-coded as indicated in the color bar and the 
redshifts are given in the top-left corner.
These Figures were produced using {\sc ifrit}.}
\label{fig:cosmo_snaps}
\end{figure*}

For our final test, we compute the reionization of a cosmological
density field in which we account for Hubble expansion but hold the
emissivity and baryon density fields constant.  We derive the initial
conditions using the same output and the same gridding technique as
in Section~\ref{sec:fedd}, but for this test we divide the volume
into $64^3$ rather than $16^3$ computational cells so that each
computational cell spans a comoving width of $125\hkpc$.  Given that the
parent simulation begins with $256^3$ baryon particles, this implies
that, on average, 64 baryon particles contribute to each cell, hence
systematic errors associated with the gridding are dominated by effects
related to low spatial resolution rather than to Poisson statistics.
We do not attempt a detailed treatment of subgrid physics as our
focus is on our radiative transfer technique (for a careful treatment
of these issues see~\citealt{mcq07}).  We evolve the ionization and
radiation fields from $z=9\rightarrow6$ with outputs spaced 1 Myr apart
assuming ($\Omega, \Lambda,H_0$) = (0.3, 0.7, 70).  We set $\nd=1$
and update the full Eddington tensor field whenever the photon number
density has changed by more than a factor of two in at least 20\% of the
volume, and after $\xhi$ drops below 0.5\% we use the optically thin
approximation (tests indicate that this yields better than 20\% accuracy 
in $\mathcal{J}$).  This computation required roughly 10000 CPU hours 
using 32 1.6 GHz Itanium2 processors in a shared-memory environment.  
Note that, although this test computation is only intended as a 
proof-of-concept for our method, it is already quite realistic in the 
sense that the gas density and emissivity fields derive from a cosmological 
hydrodynamic simulation that simultaneously reproduces a wide array of 
observations of galaxies~\citep{dav06,bou07,fin08} and the intergalactic 
medium~\citep{opp06} in the post-reionization universe.

In Figure~\ref{fig:cosmo_snaps}, we show the neutral fraction as a function 
of position in a two-dimensional slice through the computational volume at 
four representative redshifts.  Reionization proceeds in the familiar way: 
At early times, individual ionized bubbles grow around the brightest 
sources, which are strongly clustered.  As reionization proceeds, the 
individual ionized regions begin to overlap; this process can be seen to 
be well underway by $z=8$.  The volume-averaged neutral hydrogen fraction 
dips below 50\% at $z\approx7.75$.  Around this time, the mean free path of 
ionizing photons grows comparable to the length of the simulation volume 
and the ionization field becomes a network of simply-connected regions that
are largely ionized and isolated regions that are largely neutral.  As
this topology emerges, the ionizing background continues to strengthen and
the remaining neutral regions continue to shrink.  Finally, in the
post-overlap universe only regions with high recombination rates and low
emissivities remain neutral.

\begin{figure}
\centerline{
\setlength{\epsfxsize}{0.5\textwidth}
\centerline{\epsfbox{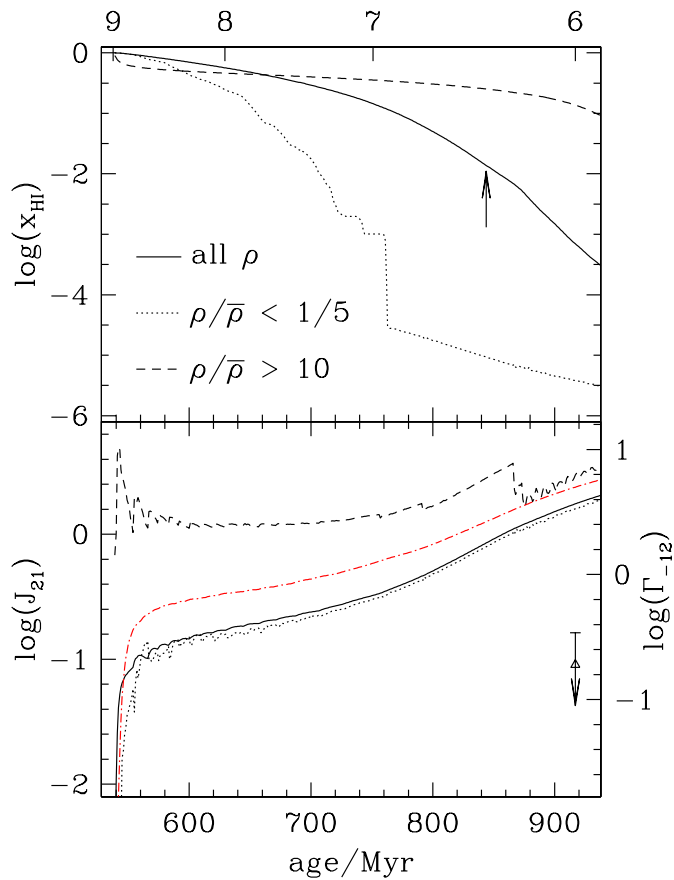}}
}
\caption{Volume-averaged neutral hydrogen fraction (top) and the ionizing
background mean intensity $J_{21}$ (bottom) as a function of the age of the 
Universe (bottom axis) and redshift (top axis).  The right axis in the bottom 
panel indicates the hydrogen ionization rate in units of $10^{-12}$ s$^{-1}$. 
In both panels, the solid line is the average over all space whereas the 
dotted and dashed lines are averaged over underdense and overdense regions, 
respectively.  The red dot-dashed curve in the bottom panel shows how the
volume-averaged $J_{21}$ varies at half our spatial resolution.  The arrow 
in the top panel indicates the observed limit on $\xhi$ at 
$z=6.4$~\citep{fan06}, while the limit in the bottom panel is 
representative of observed limits on the ionization rate at 
$z=6$~\citep{bol07}.}
\label{fig:cosmo_x_vs_t}
\end{figure}

\begin{figure}
\centerline{
\setlength{\epsfxsize}{0.5\textwidth}
\centerline{\epsfbox{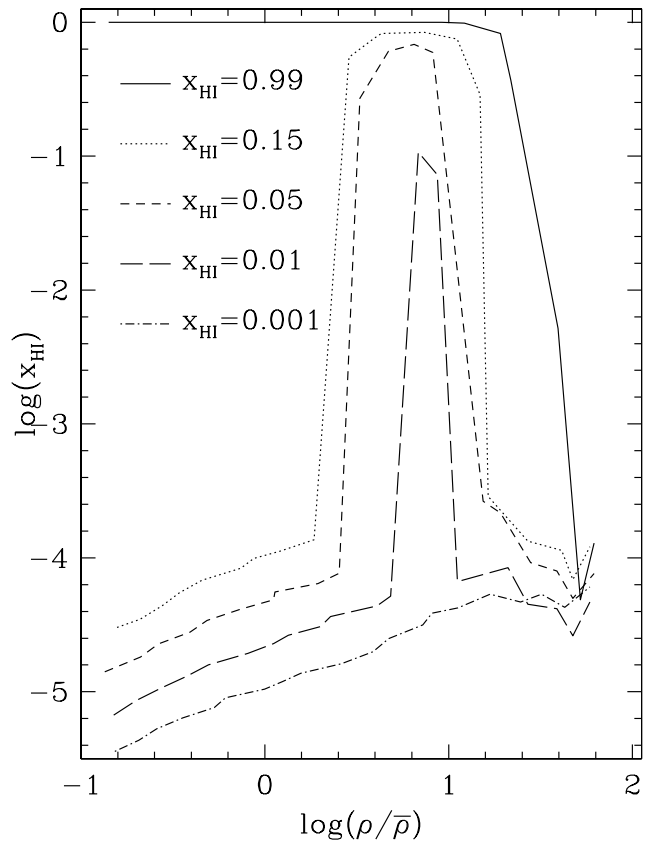}}
}
\caption{Median neutral fraction as a function of normalized density at
five Hydrogen neutral fractions as indicated.  The curves correspond to
to redshifts of (top to bottom) 8.9, 7.0, 6.7, 6.3, and 6.0 in our
simulation.}
\label{fig:cosmo_x_vs_rhonorm}
\end{figure}

In Figure~\ref{fig:cosmo_x_vs_t}, we give a more quantitative view of
how reionization proceeds in our calculation.  In the top panel, we
compare the neutral hydrogen fraction $\xhi$ averaged over the entire
computational volume with the average over underdense and overdense
regions.  At the beginning of our computation, overdense regions are
more rapidly ionized because they host the bulk of the ionizing sources.
Meanwhile, underdense regions remain more neutral because they have not 
yet been penetrated by ionization fronts.  As reionization proceeds, 
ionization fronts begin propagating into underdense regions, which 
rapidly become more highly ionized than the overdense regions.  $\xhi$ 
continues to shrink in all three density bins until $z\sim6$, by which 
point most of the universe has arrived at ionization equilibrium with 
a fairly uniform ionizing background.  At this time $\xhi$ has dropped 
to $\xhi\sim10^{-3}$, in good agreement with observations~\citep{fan06}.

We show how the relation between density and ionization fraction evolves 
in a different way in Figure~\ref{fig:cosmo_x_vs_rhonorm}.  Here, the 
different curves show the median relation at five representative ionization 
fractions as indicated.  At early times, highly overdense regions ionize to 
a neutral fraction below $<10^{-4}$ even when the cosmological mean neutral 
fraction remains at 99\% owing to their high emissivities.  Regions that are 
at and below the mean density ionize next owing to their low recombination 
rates.  Mildly overdense regions ionize last owing to their blend of 
relatively high recombination rates and low emissivities.

The reversal in the trend of ionization fraction versus overdensity that 
we find at $z\sim8$ has recently been discussed in the semi-numerical 
study of~\citet{cho08}.  In this work, it is argued that underdense 
regions should be more strongly ionized than overdense regions at late 
stages in reionization owing to their lower recombination rates.  
Our test computation, which (automatically) treats recombination rates 
realistically, supports their results while making far fewer assumptions 
regarding baryonic physics.  Unfortunately, our test cannot be used to 
study the most overdense regions before $z\approx8$ owing to the spatial
resolution of this particular simulation; this point will become clear 
when we examine the bottom panel of Figure~\ref{fig:cosmo_x_vs_t}.

In the bottom panel of Figure~\ref{fig:cosmo_x_vs_t}, we compare the time 
evolution of the volume-averaged mean specific intensity at the Lyman 
limit $J_{21} \equiv J_\nu(\nu_H)$ for the same bins of density.  We 
estimate $J_{21}$ following~\citet{mes08} by assuming that the ionizing 
photons are distributed as a power law of the form 
$J(\nu) = J_{21}(\nu/\nu_H)^{-\alpha} \times 10^{-21}$ erg s$^{-1}$ 
Hz$^{-1}$ cm$^{-2}$ sr$^{-1}$ with $\alpha=4.7$, which is appropriate 
for the typical age and metallicity of stars at $z=9$ in our simulations.  
Throughout reionization, overdense regions see a more intense ionizing 
background than the volume average.  As we saw in the top panel, this 
leads them to reionize first.  Unfortunately, this computation does not 
resolve reionization around the most overdense cells at $z > 8$ because, 
in these regions, $J_{21}$ is not smooth enough spatially.  The strong 
peaks in $J_{21}$ around the brightest sources lead to inaccurately 
computed fluxes, which in turn lead to overestimates in $J_{21}$.  
After $z=8$, however, the \hii\ regions around these cells grow and 
$J_{21}$ becomes smoother and consequently more accurate. By contrast, 
the ionizing background in underdense regions tracks the volume average 
at all times, supporting the idea that underdense regions remain more 
highly ionized at late times owing to their lower recombination rates 
rather than a stronger $J_{21}$.

The right axis converts $J_{21}$ into its associated hydrogen ionization
rate $\Gamma_{-12}\equiv\Gamma_{\mathrm{H\,I}}/10^{-12}$.  The open
triangle indicates the observed upper limit on $\Gamma_{-12}$ at $z=6$
from~\citet{bol07}.  The arrow's length combines the uncertainties in
cosmology, the observed Lyman-$\alpha$ forest effective optical depths
used to derive $\Gamma_{-12}$, and the thermal state of the intergalactic
medium.  This observation is representative of other constraints from
the literature.  Comparing the~\citet{bol07} constraint against our
simulation suggests that the simulated $\Gamma_{-12}$ is a factor of
$\sim20$ too high.  This is surprising for several reasons.  First, our
emissivity field corresponds to the galaxy population at $z=9$ and does
not account for the rise in the star formation rate density that occurs
between $z=9$ and $z=6$~\citep{opp06}.  Second, we assume an ionizing
escape fraction of 10\%, whereas~\citet{bol07} suggest that an escape
fraction of up to 20\% may be required in order to maintain the observed
ionization state of the intergalactic medium at $z=6$.  Third, we start
our test at $z=9$ whereas observations from WMAP-5 are best fit by an
instaneous reionization redshift of $11.0\pm1.4$~\citep{dun08}, which
suggests that reionization was already well underway by $z=9$.  Finally,
we do not include active galactic nuclei.  All of our assumptions,
therefore, tend to \emph{underestimate} the value of $\Gamma_{-12}$.
The source of the discrepancy could lie in cosmic variance, low spatial
resolution, or the uncertainty in the choice of ionizing escape fraction.
As a quick test, we have plotted the evolution of $J_{21}$ in a separate
test in which the same density field was divided into $32^3$ rather than
$64^3$ cells (red dot-dashed curve).  The resulting curve suggests that
some, but not all, of the discrepancy owes to poor spatial resolution.
However, further investigation into this discrepancy is beyond the scope
of this work.

\section{Summary} \label{sec:summary}

In this paper, we introduced a method that accurately and efficiently
computes continuum radiative transfer in static density fields.  The code
uses a moment-based approach to solve the equation of comoving radiative
transfer, with the Eddington tensors obtained using a long characteristics
method.  We compared several techniques for computing the Eddington
tensors that are needed to close the moment hierarchy and demonstrated
that, of the three methods that we investigated, only the method of long
characteristics has the ability to compute highly inhomogeneous radiation
fields without introducing numerical artifacts.  We found through direct 
measurement that the computation times for our long characteristics and 
moments modules scale with the number of computational cells $N_{\rm{cells}}$ 
as $N_{\rm{cells}}^{1.5}$ and $N_{\rm{cells}}^{1.0}$, respectively.  Next, 
we introduced a hybrid method for computing the evolution of nonequilibrium 
ionization fields and demonstrated that it is accurate to 1\% throughout 
our computational domain.  We combined this with our radiative transfer 
code via an efficient iterative algorithm.  The final code is regulated by 
a number of parameters, and we characterized how these parameters impact 
computation time and accuracy using a suite of low-resolution convergence 
tests.

We subjected our method to a number of standard problems in continuum
radiative transfer.  First, we verified that our code accurately computes
the growth of an \hii\ region about a source in the classical (static) case
as well as in the case of an expanding medium.  We found that, in both
cases, the radius of the resulting ionized region agrees with analytic
expectations to within the computation's spatial resolution at all times.
Next, we tested whether our code is able to produce shadows behind opaque
regions.  In agreement with previous work, we found that the moment method
introduces a small diffusion into the shadowed region~\citep{hay03}.
Nevertheless, we found that the strength of the radiation field in the
shadowed region was up to only 1\% of the value in the unshadowed region.
Finally, we computed the reionization of an expanding cosmological volume
and found qualitative agreement with other work in the literature.

Our code currently accounts for radiative and collisional ionization of 
hydrogen and helium as well as radiative recombination.  It does not 
account for recombination radiation.  Additionally, it does not follow 
the evolution of the temperature field, hence it does not account for 
photoionization suppression of star formation in low-mass halos, 
photoionization heating, recombination cooling, or shock formation.

In the future we plan to expand on our code in several ways.  First,
we have found that the computation time increases dramatically as
the universe becomes optically thin because the LC line integrals
traverse more cells before terminating either because they arrive at
the source or because they reach $\taumax$.  However, in this regime,
the optically-thin approximation, which is roughly ten times faster than
LC, becomes increasingly valid.  For this reason, we plan to study how to
transition smoothly from LC to the optically thin approximation without
introducing accuracy errors.  Second, we will generalize our method
to multifrequency radiative transfer, which is necessary for studying,
for example, ionization front hardening or \heii\ reionization.
Finally, we plan to merge our radiative transfer scheme with our version
of the cosmological galaxy formation code {\sc GADGET-2}.

\section*{Acknowledgements}

We thank Dimitrios Psaltis, Philip Pinto, Marc Metchnik, Daniel
Eisenstein, Ivan Hubeny, Moysey Brio, Rainer Wehrse, and Ben Oppenheimer
for assistance and useful discussions.  We thank Nick Gnedin for making
his excellent visualization package {\sc ifrit} publically available.
Our cosmological hydrodynamic simulation was run on the Xeon Linux
Supercluster at the National Center for Supercomputing Applications, and
many of our radiative transfer computations were run on the University of
Arizona's SGI Altix 4700 computer ``Marin."  KF acknowledges support
from a National Science Foundation Graduate Research Fellowship.
F\"O acknowledges support from NSF grant AST 07-08640.  Support for this
work was provided by the NASA Astrophysics Theory Program through grant
NNG06GH98G, as well as through grant number HST-AR-10647 from the SPACE
TELESCOPE SCIENCE INSTITUTE, which is operated by AURA, Inc. under NASA
contract NAS5-26555.  Support for this work, part of the Spitzer Space
Telescope Theoretical Research Program, was also provided by NASA through
a contract issued by the Jet Propulsion Laboratory, California Institute
of Technology under a contract with NASA.

\twocolumn

\end{document}